\documentclass[a4paper,11pt]{article}
\pdfoutput=1 

\usepackage{jheppub} 

\usepackage[T1]{fontenc} 

\usepackage{color,bm}

\newcommand{\bea} {\begin{eqnarray}}
\newcommand{\eea} {\end{eqnarray}}

\newcommand{\beq} {\begin{equation}}
\newcommand{\eeq} {\end{equation}}

\title{\boldmath Direct Production of Light Scalar in the Type-I Two-Higgs-Doublet Model   at the Lifetime Frontier of LHC}

\author[a,*]{Wei Liu,}
\author[b,\dagger]{Lei Wang,}
\author[c,\ddagger]{and Yu Zhang}

\footnotetext{$*$wei.liu@njust.edu.cn}
\footnotetext{$^\dagger$Corresponding author: leiwang@ytu.edu.cn}

\footnotetext{$^\ddagger$Corresponding author: dayu@hfut.edu.cn}

\affiliation[a]{Department of Applied Physics and MIIT Key Laboratory of Semiconductor Microstructure\\ and Quantum Sensing, Nanjing University of Science and Technology, Nanjing 210094, China}
\affiliation[b]{Department of Physics, Yantai University, Yantai
264005, P. R. China}
\affiliation[c]{School of Physics, Hefei University of Technology, Hefei 230601,China}

\abstract{A light pseudoscalar $A$  in the sufficient large $\tan\beta$ region of type-I two-Higgs-doublet model (2HDM) can be naturally a long-lived particle (LLP). We focus on  $H^{\pm}A$, $HA$ and $AA$ pair productions via the electroweak processes mediated by the bosons at the LHC, including $pp \rightarrow W^\pm/Z \rightarrow H^{\pm}/H A$  and $pp \rightarrow h \rightarrow AA$ at the 14 TeV LHC.
The possibility of probing $A$ as a LLP at the FASER-2, FACET, MoEDAL-MAPP-2, MATHUSLA is discussed. 
We find that FASER-2 fails to probe any parameter space within 0.2 GeV $< m_A <$ 10 GeV for all the considered processes. 
For 130 $< m_{H\pm} = m_H <$ 400 GeV, FACET, MoEDAL-MAPP-2 and MATHUSLA can probe $\tan \beta \lesssim 10^{4-6}$ for $m_A \lesssim 3$ GeV, and $\tan \beta \lesssim 10^{6-8}$ for 3 GeV $\lesssim m_A  <$ 10 GeV from $pp \rightarrow W^\pm/Z \rightarrow H^{\pm}/Z A$ processes. And $pp \rightarrow h \rightarrow AA$ process covers similar parameter space. All processes can surpass the current limits. }

\begin{document} 
\maketitle
\flushbottom

\section{Introduction}
\label{sec:intro}

After the discovery of the Higgs boson at the LHC~\cite{ATLAS:2012yve, CMS:2012qbp},  searching for extra Higgs becomes an important task. 
In contrast to the main detectors at the the Large Hadron Collider (LHC), such as ATLAS and CMS, which often focus on searching for heavy resonance states with the mass larger than several hundreds GeV or even few TeV, there has been a growing interest in lighter feebly interacting particles with the mass of $\mathcal{O}$(GeV). These particles typically possess longer lifetimes, enabling them to escape the main detector. 

Aiming for detecting long-lived particles~(LLPs), a series proposal of dedicated detectors based on the LHC have been put forward, including FASER~\cite{Feng:2017uoz, FASER:2018eoc, FASER:2023tle}, FACET~\cite{Cerci:2021nlb}, MoEDAL-MAPP~(MAPP)~\cite{Pinfold:2019nqj}, MATHUSLA~\cite{Chou:2016lxi, Curtin:2018mvb}, ANUBIS~\cite{Bauer:2019vqk}, Codex-b~\cite{Gligorov:2017nwh}, AL3X~\cite{Gligorov:2018vkc}, etc. Among them, FASER have already been installed and performing searches starting at the Run 3 of the LHC. These detectors have shown ample potential to probe LLPs including  dark matter~\cite{Tucker-Smith:2001myb, Dienes:2011ja, Dienes:2011sa, Dienes:2012jb, Hochberg:2015vrg}, hidden valley~\cite{Strassler:2006im, Strassler:2006ri}, dark photon~\cite{Holdom:1985ag, Bauer:2018onh, Fabbrichesi:2020wbt, Caputo:2021eaa}, axion-like particles~\cite{Peccei:1977hh, Peccei:1977ur, Jaeckel:2010ni, Bauer:2017ris}, heavy neutrinos~\cite{Gell-Mann:1979vob, Mohapatra:1979ia, Schechter:1980gr, Asaka:2005pn, Kersten:2007vk, Drewes:2015iva, Deppisch:2018eth, Amrith:2018yfb, Deppisch:2019kvs, Liu:2022kid, Liu:2022ugx, Liu:2023nxi, Zhang:2023nxy, Barducci:2023hzo, Deppisch:2023sga,Li:2023dbs, Liu:2023klu}, and vector-like lepton \cite{Bandyopadhyay:2023joz,Cao:2023smj}, cf review in Ref.~\cite{Alimena:2019zri}. Light scalars can also been potentially discovered at these detectors~\cite{Kling:2022uzy, Liu:2022nvk,Haisch:2023rqs}.

From the theory side, many models include  extended scalar sector, e.g. the popular two-Higgs-doublet model~(2HDM). The 2HDM model \cite{Lee:1973iz} is a simple extension of SM by
adding a second $SU(2)_L$ Higgs doublet, which has very rich
phenomenology, including two neutral $CP$-even Higgs bosons $h$ and
$H$, one neutral pseudoscalar $A$, and a pair of charged Higgs $H^{\pm}$. Reviews on
2HDMs and existing constraints on them can be found in Refs. \cite{Branco:2011iw,Wang:2022yhm}.


In this paper, we will focus on the type-I 2HDM, in which the long-lived scalar $A$
can be searched at FASER-2, FACET, MAPP-2 and MATHUSLA.
Its couplings with the fermion pairs are proportional to the ratio of the two vacuum expectation values, $1/\tan\beta$ which can be tiny. Given it is light, $A$ can be long-lived. 
In Ref. \cite{Kling:2022uzy} the authors studied FASER and FASER-2 reaches of $A$ as a LLP which is produced from meson decays. The final states are flying in a very forward direction, where FASER is located, hence have shown complimentary sensitivity to $A$. In this paper, we change the strategy, by searching for the $A$ via mediated~(off-shell) bosons, such as $pp \rightarrow W^\pm/Z \rightarrow H^{\pm}/H A$ and $pp \rightarrow h \rightarrow A A$. The decay products are expected to distributed in a more transverse direction, hence not only the forward detectors including FASER-2 and FACET, but also MAPP-2, MATHUSLA can potentially probe $A$.

This paper is orangised as follows, in section~\ref{sec:model}, we briefly introduce the type-I of 2HDM model. The relevant theoretical and experimental constraints on the mass of the $A$, $m_A$ and $\tan \beta$, is summarised in section~\ref{sec:currentlimits}. In section~\ref{sec:detect}, we show the sensitivity of the same parameter space by searching for $A$ at LLP detectors. Finally, the conclusion is drawn in section~\ref{sec:conclu}.

\section{Type-I of two-Higgs-doublet model}
\label{sec:model}
In the type-I 2HDM, the Higgs potential with a soft $Z_2$ symmetry breaking
can be written as 
\begin{eqnarray} \label{V2HDM} \mathcal{V}_{tree} &=& m_{11}^2
(\Phi_1^{\dagger} \Phi_1) + m_{22}^2 (\Phi_2^{\dagger}
\Phi_2) - \left[m_{12}^2 (\Phi_1^{\dagger} \Phi_2 + \rm h.c.)\right]\nonumber \\
&&+ \frac{\lambda_1}{2}  (\Phi_1^{\dagger} \Phi_1)^2 +
\frac{\lambda_2}{2} (\Phi_2^{\dagger} \Phi_2)^2 + \lambda_3
(\Phi_1^{\dagger} \Phi_1)(\Phi_2^{\dagger} \Phi_2) + \lambda_4
(\Phi_1^{\dagger}
\Phi_2)(\Phi_2^{\dagger} \Phi_1) \nonumber \\
&&+ \left[\frac{\lambda_5}{2} (\Phi_1^{\dagger} \Phi_2)^2 + \rm
h.c.\right].
\end{eqnarray}
We consider a $CP$-conservation case in which all $\lambda_i$ and
$m_{12}^2$ are real. The two complex Higgs doublet fields $\Phi_1$ and $\Phi_2$ have hypercharge $Y = 1$ and are expanded as
\begin{equation}
\Phi_1=\left(\begin{array}{c} \phi_1^+ \\
\frac{1}{\sqrt{2}}\,(v_1+\phi_1+ia_1)
\end{array}\right)\,, \ \ \
\Phi_2=\left(\begin{array}{c} \phi_2^+ \\
\frac{1}{\sqrt{2}}\,(v_2+\phi_2+ia_2)
\end{array}\right),
\end{equation}
with $v_1$ and $v_2$ being the electroweak vacuum expectation values (VEVs) and $v^2 = v^2_1 + v^2_2 = (246~\rm GeV)^2$. 
We define the ratio of the two VEVs as $\tan\beta \equiv v_2 /v_1$. 
 After spontaneous electroweak symmetry breaking, the mass eigenstates are obtained from the original fields by the rotation matrices,
\begin{eqnarray}
\left(\begin{array}{c}H \\ h \end{array}\right) =  \left(\begin{array}{cc}\cos\alpha & \sin\alpha \\ -\sin\alpha & \cos\alpha \end{array}\right)  \left(\begin{array}{c} \phi_1 \\ \phi_2 \end{array}\right) , \\
\left(\begin{array}{c}G^0 \\ A \end{array}\right) =  \left(\begin{array}{cc}\cos\beta & \sin\beta \\ -\sin\beta & \cos\beta \end{array}\right)  \left(\begin{array}{c} a_1 \\ a_2 \end{array}\right) , \\
\left(\begin{array}{c}G^{\pm} \\ H^{\pm} \end{array}\right) =  \left(\begin{array}{cc}\cos\beta & \sin\beta \\ -\sin\beta & \cos\beta \end{array}\right)  \left(\begin{array}{c} \phi^{\pm}_1 \\ \phi^{\pm}_2 \end{array}\right).
\end{eqnarray}
The $G^0$ and $G^\pm$ are Goldstones which are eaten by gauge bosons $Z$ and $W^\pm$. 
The remaining physical states are two neutral
$CP$-even states $h$, $H$, one neutral pseudoscalar $A$, and a pair of charged
scalars $H^{\pm}$. Taking the Higgs masses as the input parameters, the coupling constants
in the Higgs potential are expressed by
\begin{eqnarray}\label{poten-cba}
 &&v^2 \lambda_1  = \frac{m_H^2 c_\alpha^2 + m_h^2 s_\alpha^2 - m_{12}^2 t_\beta}{ c_\beta^2}, \ \ \ 
v^2 \lambda_2 = \frac{m_H^2 s_\alpha^2 + m_h^2 c_\alpha^2 - m_{12}^2 t_\beta^{-1}}{s_\beta^2},  \nonumber \\  
&&v^2 \lambda_3 =  \frac{(m_H^2-m_h^2) s_\alpha c_\alpha + 2 m_{H^{\pm}}^2 s_\beta c_\beta - m_{12}^2}{ s_\beta c_\beta }, \ \ \ 
v^2 \lambda_4 = \frac{(m_A^2-2m_{H^{\pm}}^2) s_\beta c_\beta + m_{12}^2}{ s_\beta c_\beta },  \nonumber \\
 &&v^2 \lambda_5=  \frac{ - m_A^2 s_\beta c_\beta  + m_{12}^2}{ s_\beta c_\beta }\, , 
 \label{eq:lambdas}
\end{eqnarray}
where the abbreviation $s_\beta\equiv \sin\beta$ and $c_\beta \equiv \cos\beta$.
When $\cos(\beta-\alpha)$ is very closed to 0, we can approximately obtain the following relations,
\begin{eqnarray}
v^2 \lambda_1 &=&  m_h^2 - \frac{t_\beta^3\,(m_{12}^2 -m_H^2  s_\beta c_\beta ) }{ s_\beta^2}\,,\nonumber \\
v^2 \lambda_2 &=& m_h^2 - \frac{ (m_{12}^2 -m_H^2  s_\beta c_\beta) }{ t_\beta s_\beta^2 }\,,\nonumber\\
\label{eq:alignment2_lambda}
v^2 \lambda_3 &=&  m_h^2 + 2 m_{H^{\pm}}^2 - 2m_H^2 -  \frac{t_\beta (m_{12}^2 -m_H^2  s_\beta c_\beta)}{  s_\beta^2}\,,\nonumber\\
v^2 \lambda_4 &=&  m_A^2-  2 m_{H^{\pm}}^2 + m_H^2+  \frac{t_\beta (m_{12}^2 -m_H^2  s_\beta c_\beta)}{  s_\beta^2}\,,\nonumber\\
v^2 \lambda_5 &=&  m_H^2 - m_A^2+  \frac{t_\beta (m_{12}^2 -m_H^2  s_\beta c_\beta)}{  s_\beta^2} \,.
\label{poten-cba0}
\end{eqnarray}

The Yukawa interactions of type-I model are 
 \bea
- {\cal L} &=&Y_{u2}\,\overline{Q}_L \, \tilde{{ \Phi}}_2 \,u_R
+\,Y_{d2}\,
\overline{Q}_L\,{\Phi}_2 \, d_R\, + \, Y_{\ell 2}\,\overline{L}_L \, {\Phi}_2\,e_R+\, \mbox{h.c.}\,, \eea
where
$Q_L^T=(u_L\,,d_L)$, $L_L^T=(\nu_L\,,l_L)$,
$\widetilde\Phi_{1,2}=i\tau_2 \Phi_{1,2}^*$, and $Y_{u2}$,
$Y_{d1,d2}$ and $Y_{\ell 1,\ell 2}$ are $3 \times 3$ matrices in family
space.

The Yukawa couplings of the neutral Higgs bosons with respect to the SM are 
\bea\label{hffcoupling} &&
y_{h}^{f_i}=\left[\sin(\beta-\alpha)+\cos(\beta-\alpha)\kappa_f\right], \nonumber\\
&&y_{H}^{f_i}=\left[\cos(\beta-\alpha)-\sin(\beta-\alpha)\kappa_f\right], \nonumber\\
&&y_{A}^{f_i}=-i\kappa_f~{\rm (for~u)},~~~~y_{A}^{f_i}=i \kappa_f~{\rm (for~d,~\ell)},\nonumber\\ 
&&{\rm with}~\kappa_d=\kappa_\ell=\kappa_u\equiv 1/\tan\beta  {\rm~ for~ type-I}.\eea 
The Yukawa interactions of the charged Higgs are 
\begin{align} \label{eq:Yukawa2}
 \mathcal{L}_Y & = - \frac{\sqrt{2}}{v}\, H^+\, \Big\{\bar{u}_i \left[\kappa_d\,(V_{CKM})_{ij}~ m_{dj} P_R
 - \kappa_u\,m_{ui}~ (V_{CKM})_{ij} ~P_L\right] d_j + \kappa_\ell\,\bar{\nu} m_\ell P_R \ell
 \Big\}+h.c.,
 \end{align}
where $i,j=1,2,3$. The neutral Higgs boson couplings with the gauge bosons normalized to the
SM are 
\beq
y^{V}_h=\sin(\beta-\alpha),~~~
y^{V}_H=\cos(\beta-\alpha),\label{hvvcoupling}
\eeq
where $V=Z,~W$. The pseudoscalar $A$ has no coupling to $V$ due to $CP$-conservation.

\section{Relevant theoretical and experimental constraints}
\label{sec:currentlimits}
We take the light $CP$-even Higgs boson $h$ as the observed 125 GeV Higgs, and assume the
approximate alignment limit, namely $\mid\sin(\beta-\alpha)\mid\to 1$, which can ensure that the tree-level couplings of $h$ to the SM particles are very closed to the SM. We take $A$ as a light LLP, $Br(h\to AA)$ is restricted by the invisible Higgs decay as $Br(h\to {\rm invisible}) < 0.24$ \cite{CMS:2016dhk,ATLAS:2015gvj,ATLAS:2017nyv}. In addition, the diphoton channel provides the most precise measurement on the property of the 125 GeV Higgs, as its signal strength \cite{ParticleDataGroup:2020ssz}
\beq
\mu_{\gamma\gamma}= 1.11^{+0.10}_{-0.09}.
\eeq
In the model, the one-loop diagrams of the charged Higgs can give additional contributions to the $h\to \gamma\gamma$ decay \cite{Wang:2013sha}, and $h\to AA$ channel enhances the total width of $h$. Therefore, the diphoton signal strength can constrain the parameter space  of the model strongly.

We consider theoretical constraints of vacuum stability, perturbativity and unitarity.
To maintain the perturbativity of theory, we demand the quartic Higgs couplings to be smaller than $4\pi$.
The conditions of vacuum stability are \cite{Deshpande:1977rw}
\beq\label{vacuum-condi}
\lambda_1 > 0,~~\lambda_2 > 0,~~\lambda_3 + \sqrt{\lambda_1\lambda_2} > 0,~~\lambda_3 + \lambda_4 - \mid\lambda_5\mid +\sqrt{\lambda_1\lambda_2} > 0.
\eeq

Besides, tree-level perturbative unitarity condition ensures perturbativity of the model up
to very high scales, which demands that the amplitudes of the scalar
quartic interactions leading to $2 \to 2$ scattering processes remain below the value of $8\pi$ at
tree-level. The requirement leads to the constraints \cite{Kanemura:1993hm,Akeroyd:2000wc},
\begin{eqnarray} \label{unitarity}
|a_{\pm}|, |b_\pm|, |c_\pm|, |{ e}_\pm|, |{ f}_\pm|, |{ g}_\pm|
\,\le\, 8\pi \,
\end{eqnarray}
with 
\begin{eqnarray}
a_\pm^{} &=& \tfrac{3}{2}(\lambda_1+\lambda_2) \pm \sqrt{\tfrac{9}{4}(\lambda_1-\lambda_2)\raisebox{0.3pt}{$^2$}+(2\lambda_3+\lambda_4)^2} \,, \nonumber\\
b_\pm^{} &=& \tfrac{1}{2}(\lambda_1+\lambda_2) \pm
\sqrt{\tfrac{1}{4}(\lambda_1-\lambda_2)\raisebox{0.3pt}{$^2$}+\lambda_4^2} \,,\nonumber \\
c_\pm^{} \,&=&\, \tfrac{1}{2}(\lambda_1+\lambda_2) \pm
\sqrt{\tfrac{1}{4}(\lambda_1-\lambda_2)\raisebox{0.3pt}{$^2$}+\lambda_5^2} \,, \nonumber\\
{ e}_\pm^{} &=& \lambda_3^{} + 2 \lambda_4^{} \pm 3 \lambda_5^{} \,, \nonumber\\
{ f}_\pm^{} \,&=&\, \lambda_3^{} \pm \lambda_4^{} \,,\nonumber\\
{ g}_\pm \,&=&\, \lambda_3^{} \pm \lambda_5^{} \,.
\end{eqnarray}

The model can give additional contributions to the oblique parameters (S,T, and U) by the exchange of extra Higgs fields in the loop diagrams of gauge boson self-energy. We employ the \textbf{2HDMC} \cite{Eriksson:2009ws} package to calculate the S, T, and U, and consider the  fit results of Ref. \cite{ParticleDataGroup:2020ssz},
\beq
S=-0.01\pm 0.10,~~  T=0.03\pm 0.12,~~ U=0.02 \pm 0.11, 
\eeq
with the correlation coefficients 
\beq
\rho_{ST} = 0.92,~~  \rho_{SU} = -0.80,~~  \rho_{TU} = -0.93.
\eeq

In our calculations, $\cos(\beta-\alpha)$, $\tan\beta$, $m_H$, $m_{H^\pm}$, and $m_A$ are randomly scanned over in the following ranges,
\bea
&&0 \leq \cos(\beta-\alpha) \leq 0.04,~~10 \leq \tan\beta \leq 10^6,\nonumber\\
&&130~ {\rm GeV} \leq m_H=m_{H^\pm} \leq 400~ {\rm GeV},~~0.2~ {\rm GeV} \leq m_A \leq 10~ {\rm GeV},
\eea
where $m_H=m_{H^\pm}$ is favored by the constraints of the oblique parameters. We obtain over 7500 points accommodating  
the theoretical constraints and the oblique parameters.
According to the first expression of Eq. (\ref{poten-cba0}),  the perturbativity requirement of $\lambda_1$ favors $m^2_{12}- m^2_H s_\beta c_\beta$ $\to$ 0 for a vary large $\tan\beta$. 
However, if both $\cos(\beta-\alpha)=0$ and $m^2_{12}- m^2_H s_\beta c_\beta =$ 0 hold exactly, solving Eq.~(\ref{poten-cba0}) and the last condition of vacuum stability in Eq. (\ref{vacuum-condi}) will lead to a strong bound, 
$m_h^2+m_A^2> m_H^2$,
which is not satisfied in the chosen parameter space.

Anyway, when $\cos(\beta-\alpha)\to 0$
 and $m_{12}^2 - m^2_H c_\beta s_\beta\to 0$, but they
 are not equal to 0 simultaneously, the previous bound is not valid anymore.
This is due to the fact that $m_{12}^2 - m^2_H c_\beta s_\beta \to 0$ can be satisfied by a very large $\tan\beta$. In the meantime, it gives non-negligible corrections to $\lambda_{1,2,3,4,5}$ in Eq.~(\ref{poten-cba0}), so Eq. (\ref{vacuum-condi}) does not lead to $m_h^2+m_A^2> m_H^2$ anymore.
Therefore, the value $m^2_{12} - m^2_H s_\beta c_\beta$ need fine-tuning, and the tuning becomes more finely as $\tan\beta$ increases. Hence, we manually take $\tan \beta \leq 10^6$. 

\begin{figure}[tb]
\begin{center}
 \epsfig{file=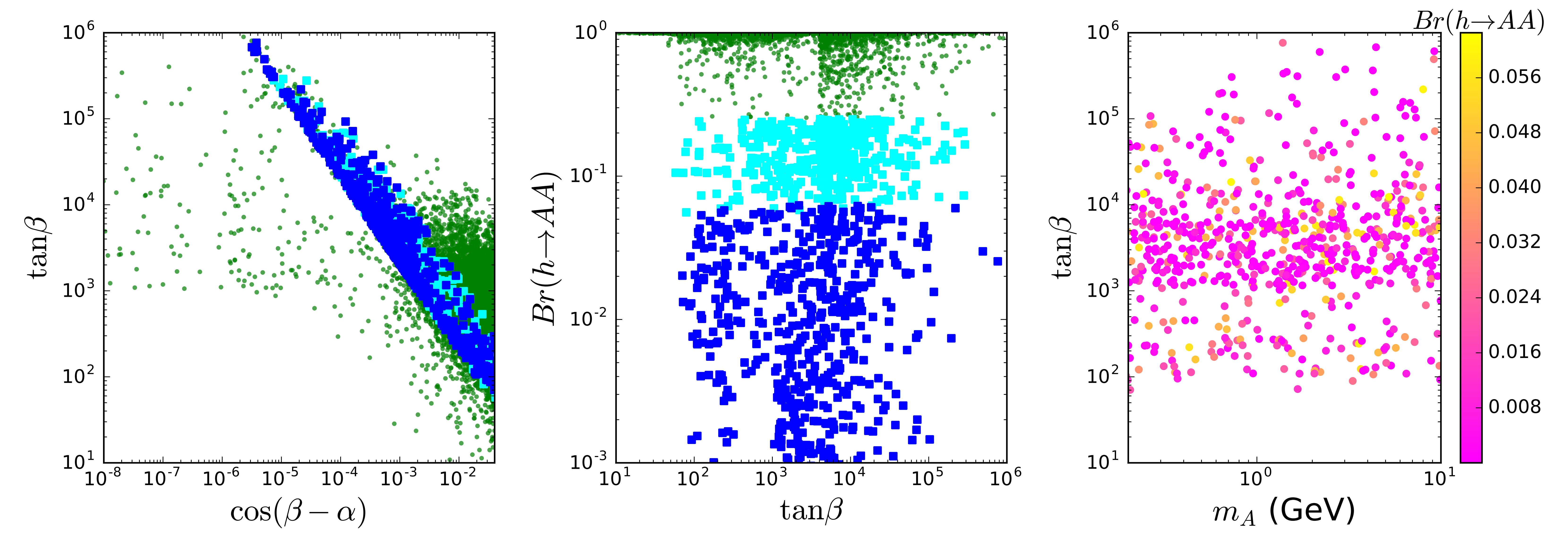,height=5.5cm}
 \end{center}
\vspace{-1.0cm} \caption{All the points are allowed by the theoretical constraints and the oblique parameters. Left and middle panels: Parameter points excluded~(green bullets) or allowed~(cyan and blue squares) by the bound $Br(h\to AA) < 0.24$.
Moreover, excluded~(cyan squares) or allowed~(blue squares) by the diphoton signal data of the 125 GeV Higgs. 
Right panels: The allowed parameter points both by the bound $Br(h\to AA) < 0.24$ and diphoton signal data on the plane of $\tan\beta$ and $m_A$, in which the varying colors indicate the value of $Br(h\to AA)$.} \label{figthe}
\end{figure}

After imposing the constraints of theory, the oblique parameters, $Br(h\to AA)<0.24$, and the diphoton signal data of the 125 GeV Higgs, we project the surviving samples in  Fig. \ref{figthe}. From it, we find that the requirement of $Br(h\to AA)<0.24$ leads to a strong correlation between $\cos(\beta-\alpha)$ and $\tan\beta$, and excludes much of the parameter space that is consistent with the theoretical constraints. 

The $h\to AA$ channel enhances the total width of $h$, and then decreases the theoretical value of $Br(h\to \gamma\gamma)$ and thus the diphoton signal strength. Therefore, combining the bound of the diphoton signal strength of the 125 GeV Higgs, $Br(h\to AA)\gtrsim 0.07$ is further excluded, which points to $\tan \beta \gtrsim 10^2$ as can be seen in Fig.~\ref{figthe}~middle. Anyway, it does not correlates to $m_A$ as shown in Fig.~\ref{figthe}~right.

Hence, in the following discussion of the sensitivity on the parameter space ($m_A$, $\tan \beta$), we further restrain it at 
\begin{align}
  0.2~{\rm GeV}\leq m_A \leq 10~{\rm GeV}, {\rm and}~ 10^2 \leq \tan \beta \leq 10^6,
\label{eq:para}
\end{align}
to satisfy the limits on the $Br(h\to AA)$. The mass of $A$ is chosen so that it can be long-lived, can be abundantly produced via mediated bosons, and is not excluded by the existing experimental limits mentioned below.

There are existing experimental constraints on a light pseudoscalar: SuperNova~\cite{Turner:1987by,Ellis:1987pk}, searches for axion-like from particles from CHARM \cite{CHARM:1985anb,Gorbunov:2021ccu}, $B$ meson decays \cite{LHCb:2015nkv,LHCb:2016awg} and $D$ meson decays \cite{LHCb:2020car} from LHCb,  Kaon decays from NA62 \cite{NA62:2021zjw}, MicroBooNE \cite{MicroBooNE:2021usw}, and E949 \cite{BNL-E949:2009dza}. These constraints will be considered in the following discussions.

\section{Detection of $A$ as a long-lived particle}
\label{sec:detect}
\begin{figure}[tb]
\begin{center}
 \includegraphics[width=0.4\textwidth]{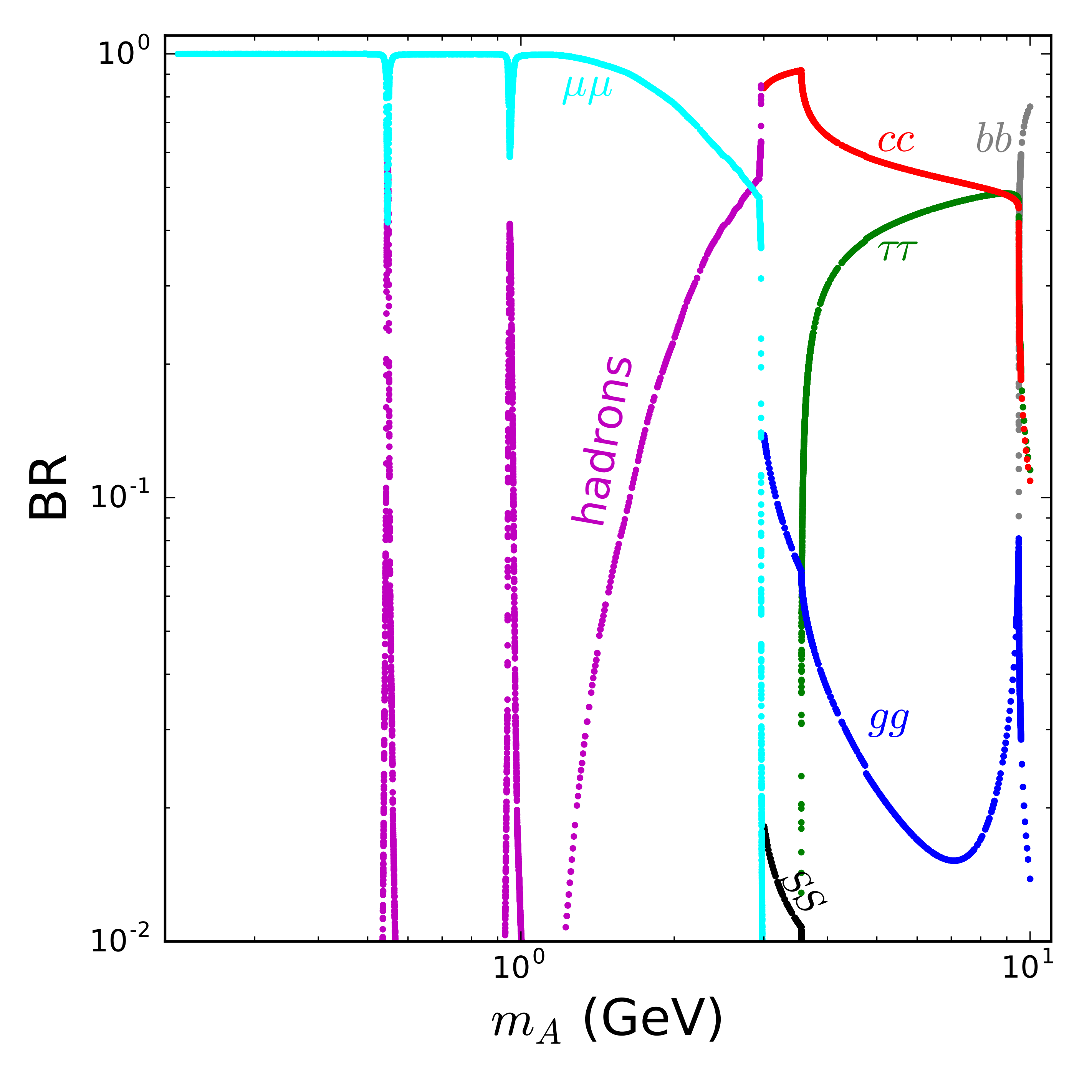}
 \includegraphics[height=0.4\textwidth,width=0.45\textwidth]{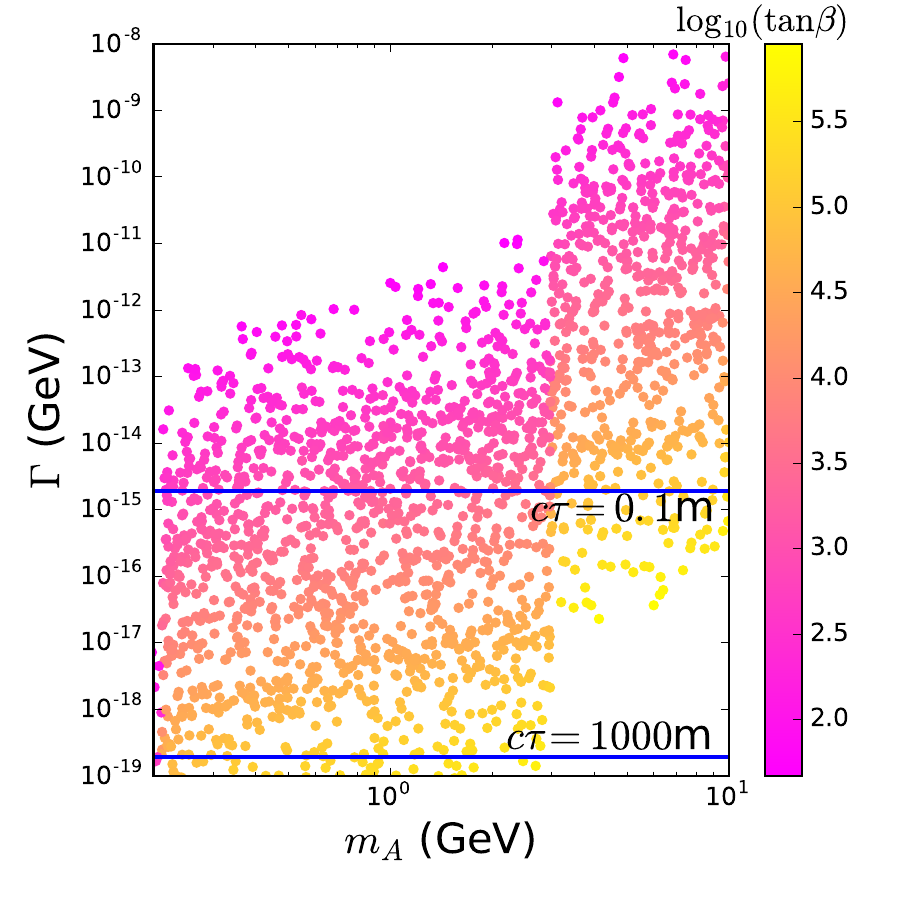}
 \end{center}
\vspace{-1.0cm} \caption{Left panel: The branching ratio of different decay channels of $A$.
Right panel: The width of $A$ versus $m_A$ satisfying the constraints of theory~\cite{Deshpande:1977rw} and the oblique parameters~\cite{Eriksson:2009ws, ParticleDataGroup:2020ssz}, $Br(h\to AA) < 0.24$~\cite{CMS:2016dhk,ATLAS:2015gvj,ATLAS:2017nyv}, and the diphoton signal data of the 125 GeV Higgs~\cite{ParticleDataGroup:2020ssz}.} \label{fig:wid}
\end{figure}

The couplings of $A$ with the fermions are proportional to $1/\tan\beta$, so a sufficiently large $\tan\beta$ can
make a light $A$ to be a LLP. Depending on its mass, the decay modes of $A$
include $A\to \ell^+\ell^-,~q\bar{q},~gg,~\gamma\gamma$.
The \textbf{2HDMC} package is employed to calculate the widths of $A\to \ell^+\ell^-,~\gamma\gamma$ for
0.2 GeV $\leq m_A \leq 10$ GeV and those of $A\to q\bar{q},~gg$ for 3 GeV $\leq m_A \leq 10$ GeV.
Below 3 GeV, the partonic description become less reliable, and we take the approaches of Ref. \cite{Kling:2022uzy,Domingo:2016yih,Holstein:2001bt} to calculate the hadronic decays of $A$. The results of the branching ratio of different decay channels of $A$ are shown in Fig.~\ref{fig:wid} left. For 0.2 GeV $< m_A \lesssim 3$ GeV, $A$ dominantly decays into muons, except the resonance of $\eta$ and $K$ mesons. For $m_A \gtrsim 3$ GeV, decays into heavy quarks and $\tau$ pairs appear and are the leading channels.

This is reflected in the total width as shown in Fig.~\ref{fig:wid} right. As shown, for 0.2 GeV $< m_A \lesssim$ 3 GeV, with $\tan \beta \gtrsim 10^4$, the total decay width becomes so small that the decay length becomes larger than order of meters. For $m_A \gtrsim 3$ GeV, since the decay into heavy quarks and $\tau$ pairs open up, the decay width increases substantially, now $\tan \beta \gtrsim 10^5$ is required to achieve order of meters decay length. And such long-lived $A$ can be captured by the LLP detectors.

The production of $A$ at the LHC is dominantly via
the following electroweak processes, with $W^\pm, Z$ bosons being off-shell:
\begin{align}
pp\to & W^{\pm} \to H^\pm A, \label{process1}\\
pp\to &       Z \to HA, \label{process2}\\
pp\to & h \to AA. \label{process5}
\end{align}

\begin{figure}[tb]
\begin{center}
\includegraphics[width=0.49\textwidth]{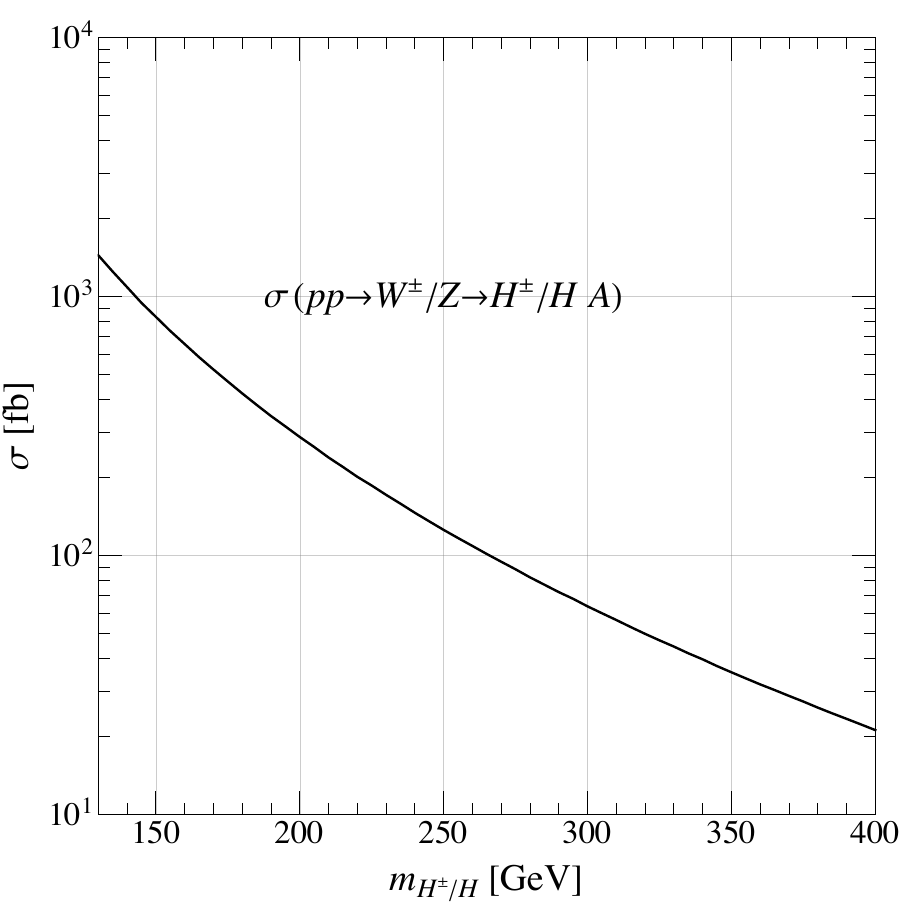}
 \end{center}
\vspace{-1.0cm} \caption{ Cross section of the process $pp \rightarrow W^\pm/Z \rightarrow H^\pm/H A$ as a function of $m_{H^\pm/H}$. We fix $m_{H^\pm} = m_H$, and $m_A =$ 1 GeV.} \label{fig:cs}
\end{figure}

For $m_H=m_{H^\pm}>>m_A$, $H\to AZ$ and 
$H^\pm\to W^\pm A$ are the dominant decay modes of 
$H$,
$H^\pm$. Hence, in the final states of $pp\to  W^{\pm}/Z \to H^\pm/H A$ processes, we have $W^{\pm}/Z + LLP$ signatures. These final states can be constrained by the mono $W/Z$ searches at the LHC, which limit $\sigma(pp \rightarrow W^\pm/Z \rightarrow H^\pm/H A) \lesssim 1.5$ pb~\cite{ATLAS:2018nda}. 
The cross section only depends on $m_{H^{\pm}/H}$ and $m_A$, since $m_{H^{\pm}/H} \gg m_{A}$ and coupling of $(W/Z) (H^{\pm}/H) A$ is $m_{W/Z}/(2v)$ for $\mid\cos(\beta-\alpha)\mid\to 1$. In Fig.~\ref{fig:cs}, we show $\sigma(pp \rightarrow W^\pm/Z \rightarrow H^\pm/H A)$ as a function of $m_{H^\pm/H}$, fixing $m_A =$ 1 GeV, at 14 TeV LHC. As $m_{H^\pm/H}$ increases, the phase space becomes smaller, so the cross section decreases rapidly from $\approx$ 1.5 pb at $m_{H^\pm/H} =$ 130 GeV, to $\approx$ 22 fb at $m_{H^\pm/H} \approx$ 400 GeV. Hence, within 130 GeV $ < m_{H^\pm/H} <$ 400 GeV, the current limits on mono $W/Z$ searches can be satisfied.

So we take two benchmark scenarios to reflect the dependence on $m_{H^{\pm}/H}$,
\begin{align}
&m_{H^{\pm}/H} = {\rm 130~GeV, }~\sigma(pp \rightarrow W^\pm/Z \rightarrow H^\pm/H A) \approx 1.5 {\rm ~pb}.\\
&m_{H^{\pm}/H} = {\rm 400~GeV, } ~\sigma(pp \rightarrow W^\pm/Z \rightarrow H^\pm/H A) \approx 22 {\rm ~fb}.
\end{align}

The production of the $A$ at the LHC is sufficiently large enough to allow for possible detection at the LLP detectors including FASER-2, FACET, MAPP-2 and MATHUSLA. Assuming the signal detection efficiency of the LLPs at these detectors is 100\%, the expected signal events $N_S$ are  controlled by the probability  $P(p, \theta, c\tau)$ of the $A$s in the final states to enter the volume of the respected detectors, such as
\begin{equation}
 N_S =   \mathcal{L}_{\text{int}}  \int \mathrm{d}p\, \mathrm{d}\theta     \frac{\mathrm{d}\sigma}{\mathrm{d}p\, \mathrm{d}\theta}  P(p, \theta, c\tau)\,,
\end{equation}
where $p, \theta$ and $c \tau$ are the three-momentum, angle between the beam line with the decay location, and proper decay length of the $A$s in the final states. 

To estimate the probability, the geometry information and location of the volume of the LLP detectors is needed. We summarise the necessary information of these detectors in the following.

\paragraph{FASER}
is placed at the very forward direction with $\theta \lesssim 10^{-3}$, and around 480 meters away from the ATLAS IP. It is already installed and collecting data sine the Run 3 of the LHC, and the first phase of its setup is aimed to accumulating 150 fb$^{-1}$ integrated luminosity~\cite{Feng:2017uoz, FASER:2018eoc, FASER:2023tle}. Since $A$ in the final states are not likely to fly in such forward direction, we do not expect the FASER to yield positive results for our considered signals, even with its second phase setup, FASER-2 at the HL-LHC with 3000 fb$^{-1}$ integrated luminosity. Nevertheless, we still list the relevant information of the FASER-2 
\begin{gather}
	L_{xy} < 1~\text{m}\,,
 ~
	 475~\text{m} < L_z < 480~\text{m} \,, ~ 
  \mathcal{L} = 3000~\text{fb}^{-1},
\end{gather} 
where $L_{xy,z}$ is the distance to the IP in $xoy$ plane or $z$ axis.
We calculate the signal events with the following procedures.

\paragraph{FACET}
is a LLP detector, recently proposed to be installed around 100 meters away from the CMS IP~\cite{Cerci:2021nlb}. Like FASER, it is a cylinder, symmetrical around the beam line in a very forward direction, but with larger solid angle coverage, $2 \times 10^{-3} \lesssim \theta \lesssim 5 \times 10^{-3}$, since it is closer to the IP, such as
\begin{gather}
	0.18~\text{m} < L_{xy} < 0.5~\text{m}\,,
 ~ 
	 101~\text{m} < L_z < 119~\text{m} \,,~ 
  \mathcal{L} = 3000~\text{fb}^{-1}.
\end{gather} 

\paragraph{MAPP}
is a new sub-detector of the MoEDAL detector, aiming for searching LLPs~\cite{Pinfold:2019nqj}. 
Unlike the FASER-2 and FACET detectors, it is a box, which is not symmetrical around the beam line.
There are two phase setup of the MAPP detectors. The first phase is already underway, in the UA83 tunnel about 100 meters away from the LHCb IP. The second phase will be installed in the UGC1 gallery roughly 50 meters away from the same IP, for the HL-LHC~\cite{MoEDAL-MAPP:2022kyr}.
Nevertheless, we only take the second phase, MAPP-2 into account to get the optimistic results. It has two polyhedrons each characterised by eight points~\cite{MoEDAL-MAPP:2022kyr},
with the first polyhedron by
\begin{gather}
 	\text{Point 1} (4.00, \ \ \ 1, -61.39), \
	\text{Point 2} (16.53, \ \  1, -35.45), \\ \nonumber
	\text{Point 3} (3.27, \ \ \ 1, -52.83), \
	\text{Point 4} (12.24, \ \  1, -33.63), \\ \nonumber
	\text{Point 5} (4.00, -2, -61.39), \
	\text{Point 6} (16.53, -2, -35.45), \\ \nonumber
	\text{Point 7} (3.27, -2, -52.83), \
	\text{Point 8} (12.24, -2, -33.63).
\end{gather} 
And the second polyhedron by
\begin{gather}
 	\text{Point 1} (16.53, \ \ \ 1, -35.45), \
	\text{Point 2} (19.00, \ \  1, -29.63), \\ \nonumber
	\text{Point 3} (12.24, \ \ \ 1, -33.63), \
	\text{Point 4} (14.57, \ \  1, -28.63), \\ \nonumber
	\text{Point 5} (16.53, -2, -35.45), \
	\text{Point 6} (19.00, -2, -29.63), \\ \nonumber
	\text{Point 7} (12.24, -2, -33.63), \
	\text{Point 8} (14.57, -2, -28.63).
\end{gather} 
These points represents the distance~(in meters) to the IP of LHCb in $x, y$ and $z$ axis, respectively.

\paragraph{MATHUSLA}
is a large surface detector on the ground of the CMS IP~\cite{Chou:2016lxi, Curtin:2018mvb, MATHUSLA:2019qpy}. Unlike the aforementioned detectors, it is placed at the transverse direction, with the largest solid angle coverage, proposed to be installed at the HL-LHC era,
\begin{gather}
	-100~\text{m} < L_{x} <100~\text{m}\,,
 ~100~\text{m} < L_{y} <120~\text{m}\,,~ 
	 100~\text{m} < L_z < 300~\text{m} \,,~ 
  \mathcal{L} = 3000~\text{fb}^{-1}.
\end{gather} 
Although the installation of the MATHUSLA is still in discussion, we take it as an example to show the most positive outcome of the LLP detectors.

For the four detectors, given whether they are symmetrical around the beam line,
we use two strategies to estimate the probability of the $A$ entering into the detector volume. FASER-2 and FACET have cylindrical symmetry around the beam line, hence
\begin{equation}
 P(p,\theta,c\tau) =
 \left( e^{-(L-\Delta)/d }- e^{-L/d} \right) \Theta(R-\tan\theta L) \approx \frac{\Delta}{d} e^{-L/d}  \Theta(R- \theta L) \
\end{equation}
where $\Theta$ is the Heaviside function, $p$ is the three momentum of $A$, $d = c \tau p/m_A$  is the decay length of $A$ in the lab frame, $\theta$ is the angle of the momentum of $A$ to the beam line, $L$ is the farthest longitudinal distance to the IP of the detector, $\Delta$ is the longitudinal length of the detector, and $R$ is the radius in the $xoy$ plane, respectively.

The differential cross section $\frac{d \sigma}{dp d \theta}$ needs to be simulated via Monte-Carlo methods. The processes are generated after feeding the Universal FeynRules Output~(UFO)~\cite{Alloul:2013bka, Degrande:2011ua} to the event generator {\tt MadGraph5aMC$@$NLO} v3.5.1~\cite{Alwall:2014hca}. 
We generate $10^4$ events for MAPP-2 and MATHUSLA. In order to increase the statistics in FASER-2 and FACET-2, we generate $10^5$ events and further require $\eta_{A} >$ 2.5.
After that, the differential cross section is obtained by the event generator, whereas the effects of shower, hadronization, etc are handled by {\tt PYTHIA v8.306}~\cite{Sjostrand:2014zea}. 

Nonetheless, MAPP-2 and MATHUSLA do not possess cylindrical symmetry, and the probability in such detectors are not analytical. Hence, we use full Monte-Carlo methods, as we further use inverse sampling methods to simulate the exponential distribution of the lab decay length of the $A$.

\begin{figure}[tb]
\begin{center}
 \includegraphics[width=0.49\textwidth]{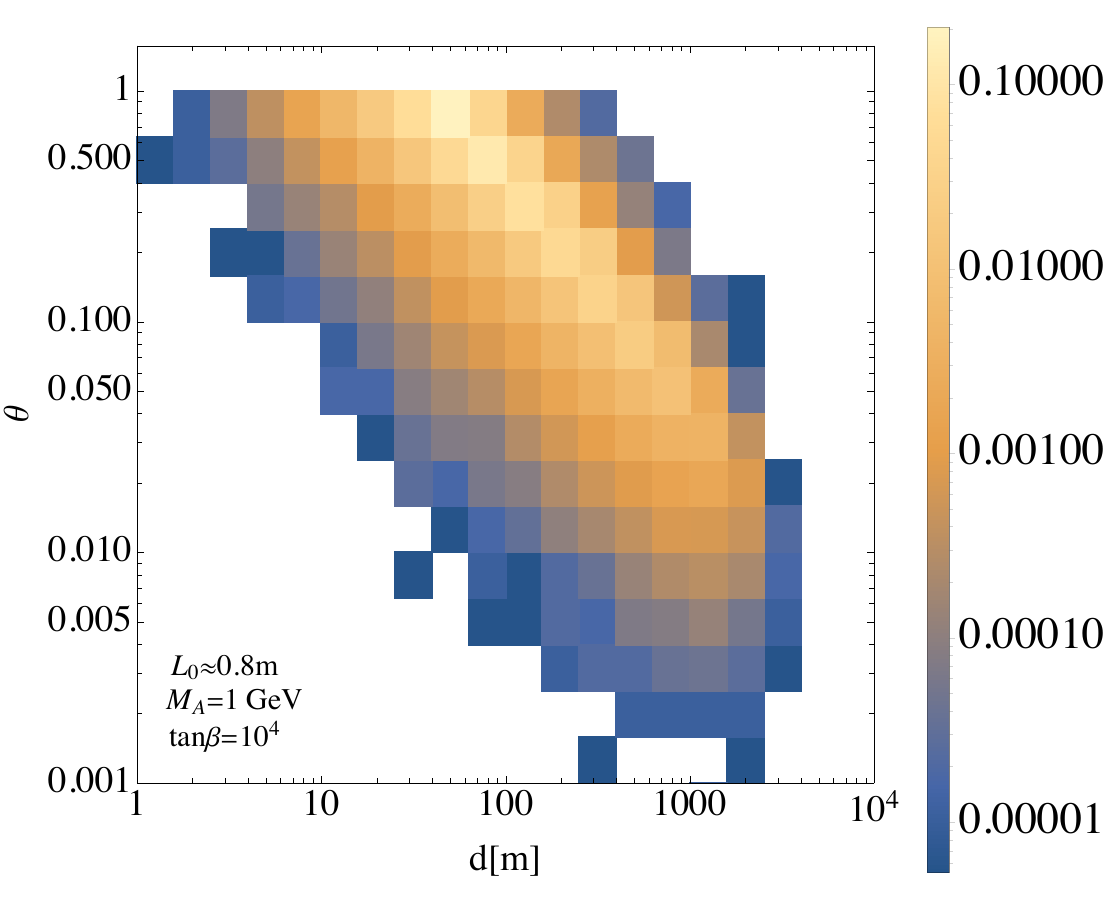}
  \includegraphics[width=0.49\textwidth]{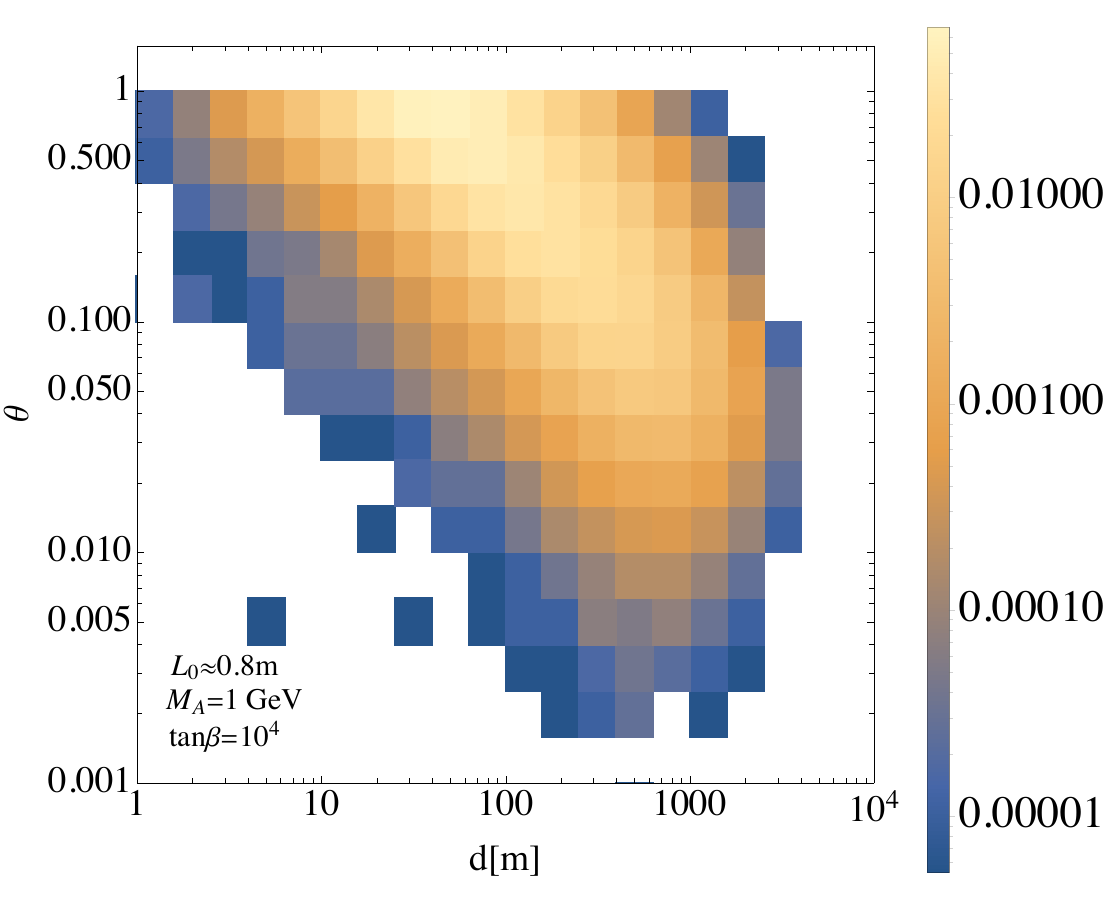}
 \end{center}
\vspace{-1.0cm} \caption{Two dimensional histograms for the ($d$, $\theta$) of $A$ from the processes $pp\to W^{\pm}/Z \to H^\pm/H A$~(left), and $pp \rightarrow h \rightarrow AA$~(right) at the 14 TeV LHC. $d = c \tau p /m_A$ is the decay length of $A$ in the lab frame, and $\theta$ is the angle of the momentum of $A$ to the beam line.
The probability distribution function is illustrated by the color, which is normalized to one.
We fix $m_A=$ 1 GeV $\tan \beta = 10^4$ so $c \tau \approx$ 0.8 m. The masses of heavy and charged Higgs is fixed as $m_{H^{\pm}/H} = 130$ GeV.} \label{fig:dis}
\end{figure}

To graphically illustrate  $P(p,\theta,c\tau)$ for different detectors and processes, we show the ($d$, $\theta$) distribution of $A$ from the processes $pp\to W^{\pm}/Z \to H^\pm/H A$~(left), and $pp \rightarrow h \rightarrow AA$~(right) at the 14 TeV LHC, in Fig.~\ref{fig:dis}. We fix $m_A=$ 1 GeV, $\tan \beta = 10^4$ so $c \tau \approx$ 0.8 m. The masses of heavy and charged Higgs are fixed as $m_{H^{\pm}/H} = 130$ GeV. The distribution roughly follows, $p \theta = p_T \approx m_{W,Z,H}/2 \approx 50 $ GeV. Hence, for FASER-2 and FACET, $\theta \lesssim 10^{-(2-3)}$ corresponds to $p \gtrsim 10^{4-5}$ GeV, which roughly exceeds the collision energy of the LHC. So $A$ rarely enters into the volume of the FACET and FASER-2. In the meantime, MAPP-2 and MATHUSLA also have larger detector volume, so we expect better sensitivity at MAPP-2 and MATHUSLA.

To derive the sensitivity, estimating the background is necessary. For such LLPs signal, the SM background contains decays of neutral hadrons, punch-through muons, neutrino interaction with the detectors, etc~\cite{FASER:2023tle}. 
For FASER, the first search for LLPs have been performed by looking for dark photons~\cite{FASER:2023tle}. The total background is shown to be negligible, $\mathcal{O}(10^{-3})$ events for 27 fb$^{-1}$. Hence, no background is also expected at FASER-2 with 3000 fb$^{-1}$. For FACET, following Ref.~\cite{Cerci:2021nlb}, background from neutral hadrons is overwhelming for $m_A <$ 0.8 GeV. Otherwise, the background can be considered to be negligible. Since there lacks information of background for MAPP-2, the estimation is not plausible. As MAPP-2 does not contain calorimeter, so low energy background is hard to reject. Hence, there can be significant background in MAPP-2, and the estimation requires further information provided in the future. When it comes to MATHUSLA, as a surface detector, the leading background becomes the cosmic ray air showers~\cite{MATHUSLA:2018bqv}. This can be highly rejected by  using a combination of timing and trajectory reconstruction. Other background from muons and neutrinos can also be efficiently rejected.

Therefore, we take background free assumption for all detectors, except for MAPP-2.
To avoid the mass range which might suffer from large background, we further require $m_A > $ 0.8 GeV for FACET-2.

Following the above procedures, we estimate the sensitivity to the $\tan \beta$ at the FASER-2, FACET, MAPP-2 and MATHUSLA. The 95\% confidence level~(CL) sensitivity is set by requiring $N_S \gtrsim 3$, with no background assumption.
Although we set our interested parameter space in Eq.~\ref{eq:para} with $\tan \beta < 10^6$, we extend it to $\tan \beta < 10^8$ as the LLP detectors might be sensitive to. Nevertheless, such values of $\tan \beta$ requires over fine-tuning, so we put the line labelled 'theory' to reflect that~\footnote{Such $\tan \beta \gtrsim 10^6$ requires $|m_{12}^2-m_H^2 s_\beta c_\beta| \lesssim 10^{-14}$, which is over fine-tuned.}.

\begin{figure}[tb]
\begin{center}
 \includegraphics[width=0.8\textwidth]{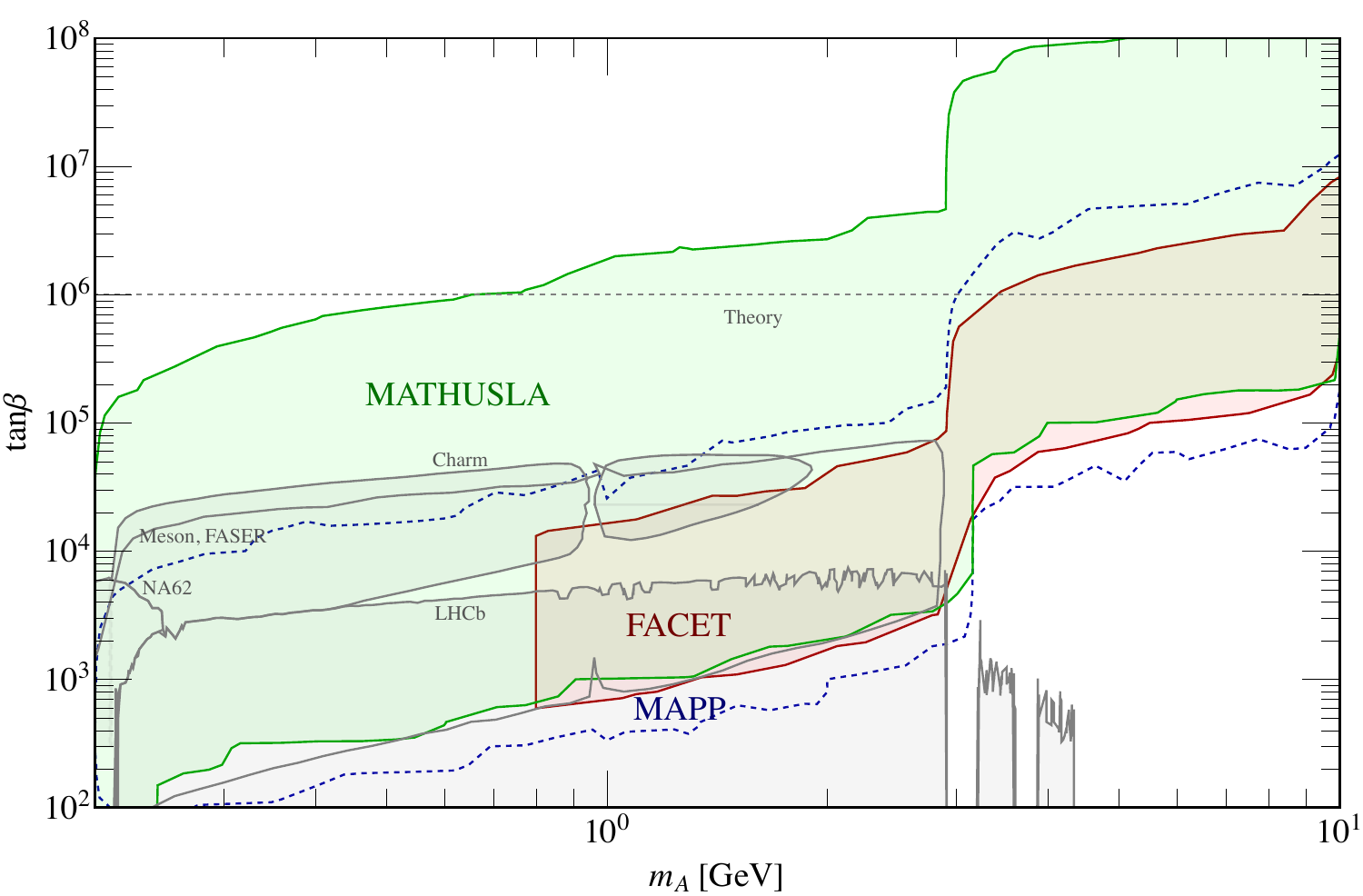}
 \end{center}
\vspace{-1.0cm} \caption{The 95\% CL sensitivity to the $\tan \beta$ as a function of $m_A$, via searching for the long-lived particle signature of the $pp \rightarrow W^\pm/Z \rightarrow H^\pm/H A$ at FACET, MAPP-2 and MATHUSLA. The region of MAPP-2 is not shaded and the boundary is dashed, since the number of background requires further study.
The current limits from CHARM~\cite{CHARM:1985anb,Gorbunov:2021ccu}, LHCb~\cite{LHCb:2015nkv, LHCb:2016awg, LHCb:2020car}, NA62~\cite{NA62:2021zjw} is overlaid for comparison. The gray curve is the expected sensitivity from searching $A$ from meson decay at FASER-2~\cite{Kling:2022uzy}. The line 'theory' is put at 
 $\tan \beta \leq 10^6$, reflects where the value of $\tan \beta$ requires over fine-tuning. We fix $m_{H^{\pm}/H} = $ 130 GeV.
} \label{fig:con}
\end{figure}

We begin with the process $pp\to W^{\pm}/Z \to H^\pm/H A$
with $m_{H^\pm/H} =$ 130 GeV, as the results are shown in Fig.~\ref{fig:con}. The current limits from CHARM~\cite{CHARM:1985anb,Gorbunov:2021ccu}, LHCb~\cite{LHCb:2015nkv, LHCb:2016awg, LHCb:2020car}, NA62~\cite{NA62:2021zjw} are overlaid for comparison. The CHARM experiments can already probe $\tan \beta \sim 5 \times 10^4$ for $m_A \lesssim$ 2 GeV, where $A$ is long-lived. Under that, via searching for the anomalous decays of various mesons, LHCb and NA62 can probe  $\tan \beta \lesssim 4 \times 10^3$ in $m_A \lesssim$ 3 GeV. Hence, there exists a valley between the three experiments, which can be filled by searching for the long-lived $A$ from the meson decays at FASER and FASER-2~\cite{Kling:2022uzy}. 

Nevertheless, for our processes, $pp \rightarrow W^\pm/Z \rightarrow H^\pm/H A$, the $A$ in the final states are not likely to distributed in such very forward direction where FASER-2 located as indicated by Fig.~\ref{fig:dis}. Subsequently, as expected, there is no sensitivity at either FASER or FASER-2. 
On the other hand, since FACET has larger solid angle coverage, it yields positive results. Although FACET is located 100 meters away, $A$ is so light that it is boosted by $\mathcal{O}(10)$ times, so the sensitivity contour roughly follows the distribution of the decay length $c \tau \sim 1$ m, as shown in Fig.~\ref{fig:wid}. 
For $m_A < 3$ GeV, FACET can touch $\tan \beta \lesssim 5 \times 10^4$, which is already ruled out by either the CHARM or meson decays at the FASER-2. Aiming at $m_A \gtrsim 3$ GeV, multiple channels for $A$ to decay into mesons open up, such that the required $\tan \beta$ to make $A$ long-lived suddenly jumps. In such mass range, $A$ cannot be generated by light meson decays, so there rarely exists any current limits, only the heavy meson decays at the LHCb giving rather softer constraints. Nevertheless, such heavy $A$ can still be produced via off-shell weak boson, and we can probe $\tan \beta  \lesssim 10^{6-7}$ at FACET, which is not explored yet.

Although MAPP-2 is placed only half distant away comparing to FACET, due to its much larger solid angle coverage, up to $\theta \lesssim 0.1$, it has shown better sensitivity, extends roughly 0.2 magnitude more both at lower and higher edge of $\tan \beta$ in a very similar mass range of $A$. However, the sensitivity can be potentially reduced, since background free assumption at MAPP-2 is not promised.
The situation becomes even better for MATHUSLA, with large coverage in angle~(up to $\theta \lesssim 1$), and so in distance, it can probe $\tan \beta  \lesssim 10^{6-7}$ for $m_A < 3$ GeV, and $\tan \beta  \lesssim 10^8$ for 3 GeV $< m_A < 10$ GeV, surpassing the current limits by at least one magnitude.

\begin{figure}[tb]
\begin{center}
 \includegraphics[width=0.8\textwidth]{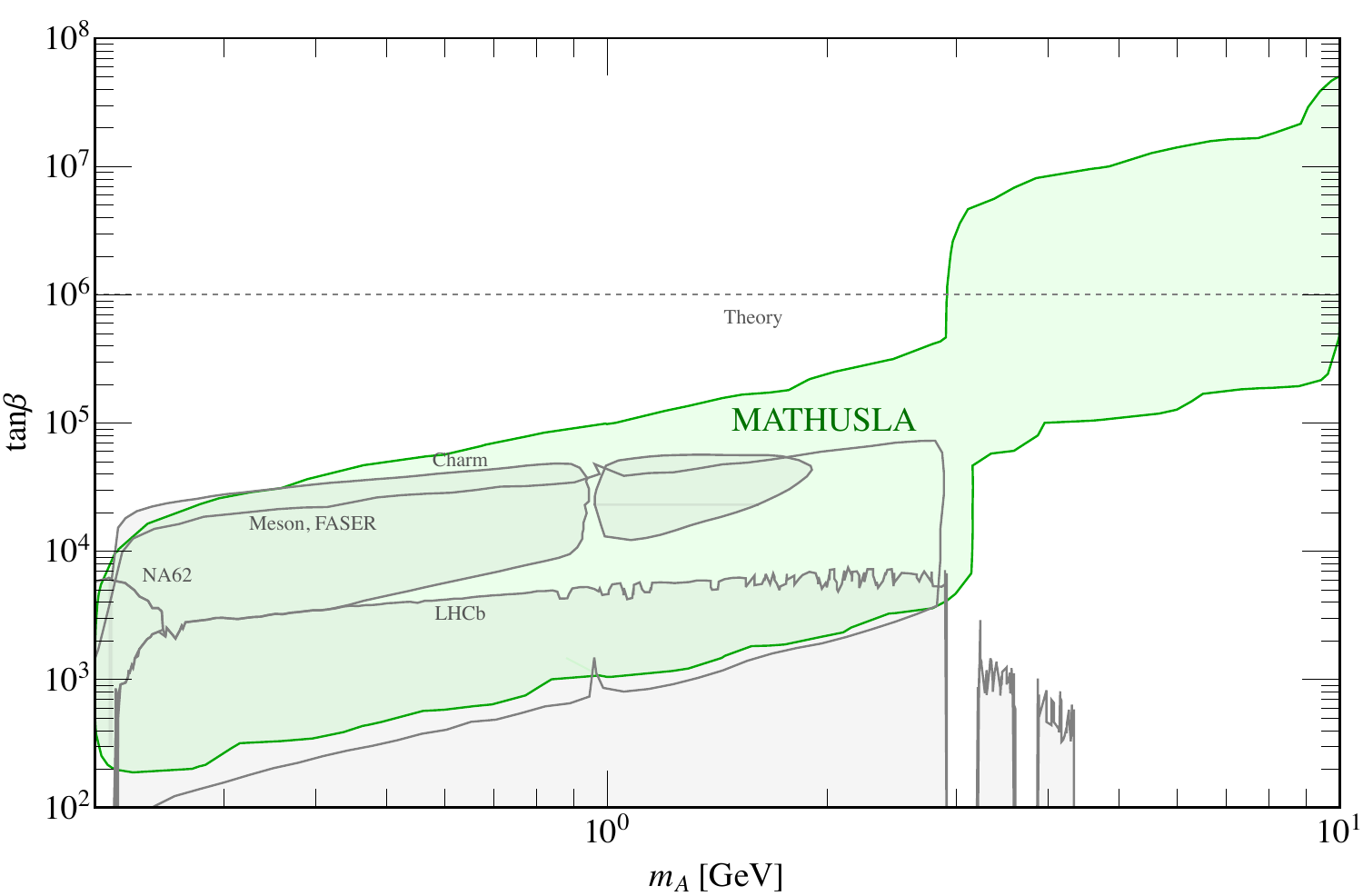}
 \end{center}
\vspace{-1.0cm} \caption{Same as Fig.~\ref{fig:con}, while $m_{H^\pm/H} = $ 400 GeV.} \label{fig:con400}
\end{figure}

It is interesting to discuss the dependence of the sensitivity on different $m_{H^\pm/H}$. In Fig.~\ref{fig:con400}, we show the sensitivity from the same processes, while the mass of $H^\pm/H$ is larger, $m_{H^\pm, H} = $ 400 GeV. As indicated in Fig.~\ref{fig:cs}, the cross section of $\sigma(pp \rightarrow W^\pm/Z \rightarrow H^\pm/H A)$ becomes about one hundred times smaller, so the sensitivity is expected to be reduced. For this benchmark, FACET no longer covers any parameter space at all, due to its low angular coverage. The sensitivity of MAPP-2 also disappears, due to the small cross section. When it comes to MATHUSLA, benefited from its large geometrical coverage, the sensitivity is still better than the current limits at CHARM, and probes new region where $\tan \beta \lesssim 10^7$ for $m_A \gtrsim 
$ 3 GeV.

As shown in Fig.~\ref{fig:cs}, $\sigma(pp \rightarrow W^\pm/Z \rightarrow H^\pm/H A)$ decreases as $m_{H^\pm/H}$ becomes larger. So the benchmark $m_{H^\pm, H} = 400$ GeV is serving to show the worst sensitivity for 130 GeV $ < m_{H^\pm/H} <$ 400 GeV. Since we have shown that with $m_{H^\pm/H} = 400$ GeV, MATHUSLA is able to obtain comparable sensitivity for  $m_A \lesssim 
$ 3 GeV to the current limits, and explores new parameter space for $m_A \gtrsim 
$ 3 GeV, thus we can conclude that for the whole parameter space of 130 GeV $ < m_{H^\pm, H} <$ 400 GeV, we are expecting better sensitivity on $(m_A, \tan \beta)$ at LLP detectors than the current limits from the processes $\sigma(pp \rightarrow W^\pm/Z \rightarrow H^\pm/H A)$.

\begin{figure}[tb]
\begin{center}
\includegraphics[width=0.8\textwidth]{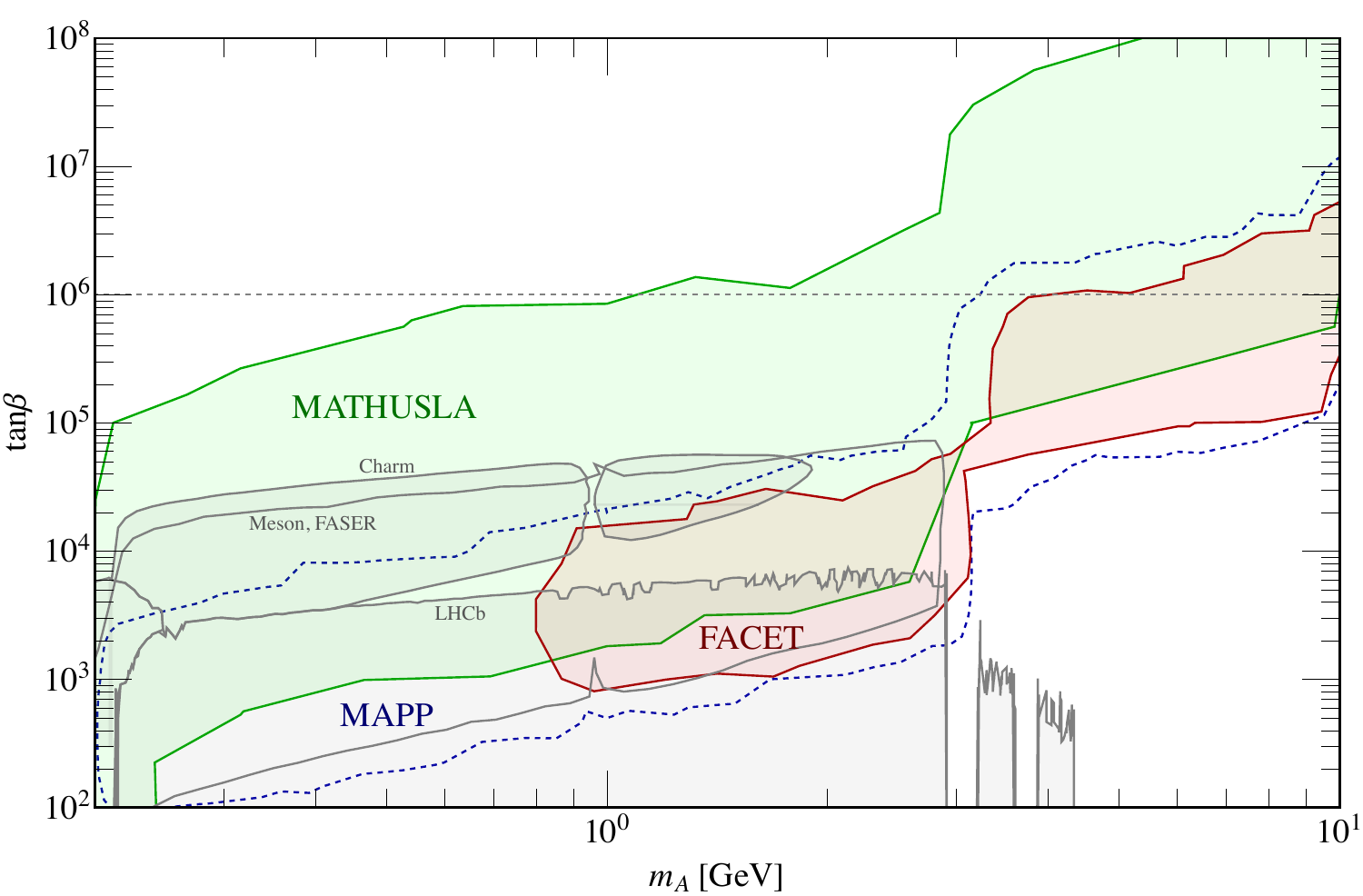}
 \end{center}
\vspace{-1.0cm} \caption{Same as Fig.~\ref{fig:con}, whereas the processes is $pp \rightarrow h \rightarrow A A$. We fix $Br(h \rightarrow A A) =$ 0.01.} \label{fig:haa}
\end{figure}

The pseudoscalar $A$ can also be pair-produced from the decay of the SM Higgs. In Fig.~\ref{figthe}, we have shown that $Br(h \rightarrow A A) = 0.01$ can be obtained in a very broad parameter space. At 14 TeV LHC, we have the full SM Higgs production, $\sigma(pp \rightarrow h) \approx 56$~pb~\cite{LHCHiggsWG}, hence $\sigma(pp \rightarrow h \rightarrow A A)$ can reach $\approx$ 0.56 pb, which is at least comparable to $\sigma(pp \rightarrow W^\pm/Z \rightarrow H^\pm/H A)$. Regarded as an optimistic prospects, we fix $Br(h \rightarrow A A) = 0.01$ and show the sensitivity from LLP searches of this process in Fig.~\ref{fig:haa}. The final states of the SM Higgs decays are more likely to distribute at the transverse direction, which FASER-2 and MAPP-2 can not cover, and the sensitivity of them becomes worse comparing to the case where $m_{H^{\pm}/H} = $ 130 GeV for the processes $pp \rightarrow W^\pm/Z \rightarrow H^{\pm}/H A$. MATHUSLA are closer to the transverse direction, so still yield similar sensitivity, strongly surpass the current limits.

\section{Conclusion}
\label{sec:conclu}

In this work, we have explored the possibility to probe the type-I 2HDM model via searching for the LLP final states of the light Higgs $A$, at FASER-2, FACET, MAPP-2 and MATHUSLA, directly produced via mediated bosons including $pp \rightarrow W^\pm/Z \rightarrow H^\pm/H A$ and $pp \rightarrow h \rightarrow A A$.  

We have focused on light $A$ with masses up to 10 GeV. The decay length of $A$, is only dependent on its masses and the ratio between the vevs of the two Higgs doublets, $\tan \beta$. When $\tan \beta$ is sufficiently large, $A$ can have decay length $\mathcal{O}$(m), resulting unique signatures as LLPs at the LHC. Such LLP signatures can be smoking-gun of the physics beyond the SM, as the SM background is nearly zero. Targetting at such signatures, multiple LLP detectors are proposed. Background free assumption can be safe for FASER-2, FACET and MATHUSLA, while MAPP-2 requires future studies. Among them, FASER is already installed and collecting data since run 3 of the LHC.

We have discussed the sensitivity reach of these detectors from the $pp \rightarrow W^\pm/Z \rightarrow H^\pm/H A$ processes. As the cross section also dependent on $m_{H^\pm}$ and $m_{H}$, we take two benchmarks where $m_{H^\pm} = m_H = 130$ and 400 GeV, respectively. The sensitivity on the parameter space $(m_A, \tan \beta)$ is obtained and compared to the current limits mainly from CHARM and LHCb, as well as the projected limits of searching for $A$ from meson decays at FASER-2 in Ref.~\cite{Kling:2022uzy}. We have shown that, since the final states from direct production are not likely to distributed in such forward direction where FASER-2 located, it has no sensitivity at all. Anyway, other LLP detectors, especially MATHUSLA, have shown at least comparable sensitivity to the current limits for $m_A \lesssim$ 3 GeV, with $\tan \beta \lesssim 10^5$, and extends the sensitivity for $m_A \gtrsim$ 3 GeV with $\tan \beta \lesssim 10^{7-8}$ where have never been probed. Giving that $\sigma(pp \rightarrow W^\pm/Z \rightarrow H^\pm/H A)$ decreases as $m_{H^\pm/H}$ increases, we can expect the same conclusion within the range 130 GeV $<m_H^\pm= m_H <$ 400 GeV.

The sensitivity from the channels of the SM Higgs decays, $pp \rightarrow h \rightarrow A A$ is also shown. Giving that $Br(h \rightarrow AA)$ is independent of both $m_A$ and $\tan \beta$, we fix it at $Br(h \rightarrow AA) = 0.01$.
For such channels, $A$s in the final state are situated at directions closer to the transverse ones, hence FASER-2 fails to show any sensitivity, and the sensitivity of FACET and MAPP-2 shrinks. Anyway,  MATHUSLA can still show comparable sensitivity just as  the case of $pp \rightarrow W^\pm/Z \rightarrow H^\pm/H A$ when $m_H^\pm = m_H =$ 130 GeV.

\acknowledgments
This work was supported by the National Natural Science Foundation
of China under grants No.12205153, 11975013, 11805001,  the Natural Science Foundation of
Shandong province ZR2023MA038, and the Fundamental Research Funds for the Central Universities under Grant No.JZ2023HGTB0222.

\bibliographystyle{JHEP}
\bibliography{submit.bib}

\providecommand{\href}[2]{#2}\begingroup\raggedright\begin{thebibliography}{10}

\bibitem{ATLAS:2012yve}
{\scshape ATLAS} collaboration, G.~Aad et~al., \emph{{Observation of a new particle in the search for the Standard Model Higgs boson with the ATLAS detector at the LHC}}, \href{http://dx.doi.org/10.1016/j.physletb.2012.08.020}{\emph{Phys. Lett. B} {\bf 716} (2012) 1--29}, [\href{http://arxiv.org/abs/1207.7214}{{\tt 1207.7214}}].

\bibitem{CMS:2012qbp}
{\scshape CMS} collaboration, S.~Chatrchyan et~al., \emph{{Observation of a New Boson at a Mass of 125 GeV with the CMS Experiment at the LHC}}, \href{http://dx.doi.org/10.1016/j.physletb.2012.08.021}{\emph{Phys. Lett. B} {\bf 716} (2012) 30--61}, [\href{http://arxiv.org/abs/1207.7235}{{\tt 1207.7235}}].

\bibitem{Feng:2017uoz}
J.~L. Feng, I.~Galon, F.~Kling and S.~Trojanowski, \emph{{ForwArd Search ExpeRiment at the LHC}}, \href{http://dx.doi.org/10.1103/PhysRevD.97.035001}{\emph{Phys. Rev. D} {\bf 97} (2018) 035001}, [\href{http://arxiv.org/abs/1708.09389}{{\tt 1708.09389}}].

\bibitem{FASER:2018eoc}
{\scshape FASER} collaboration, A.~Ariga et~al., \emph{{FASER\textquoteright{}s physics reach for long-lived particles}}, \href{http://dx.doi.org/10.1103/PhysRevD.99.095011}{\emph{Phys. Rev. D} {\bf 99} (2019) 095011}, [\href{http://arxiv.org/abs/1811.12522}{{\tt 1811.12522}}].

\bibitem{FASER:2023tle}
{\scshape FASER} collaboration, H.~Abreu et~al., \emph{{Search for dark photons with the FASER detector at the LHC}}, \href{http://dx.doi.org/10.1016/j.physletb.2023.138378}{\emph{Phys. Lett. B} {\bf 848} (2024) 138378}, [\href{http://arxiv.org/abs/2308.05587}{{\tt 2308.05587}}].

\bibitem{Cerci:2021nlb}
S.~Cerci et~al., \emph{{FACET: A new long-lived particle detector in the very forward region of the CMS experiment}}, \href{http://dx.doi.org/10.1007/JHEP06(2022)110}{\emph{JHEP} {\bf 06} (2022) 110}, [\href{http://arxiv.org/abs/2201.00019}{{\tt 2201.00019}}].

\bibitem{Pinfold:2019nqj}
J.~L. Pinfold, \emph{{The MoEDAL Experiment at the LHC\textemdash{}A Progress Report}}, \href{http://dx.doi.org/10.3390/universe5020047}{\emph{Universe} {\bf 5} (2019) 47}.

\bibitem{Chou:2016lxi}
J.~P. Chou, D.~Curtin and H.~J. Lubatti, \emph{{New Detectors to Explore the Lifetime Frontier}}, \href{http://dx.doi.org/10.1016/j.physletb.2017.01.043}{\emph{Phys. Lett. B} {\bf 767} (2017) 29--36}, [\href{http://arxiv.org/abs/1606.06298}{{\tt 1606.06298}}].

\bibitem{Curtin:2018mvb}
D.~Curtin et~al., \emph{{Long-Lived Particles at the Energy Frontier: The MATHUSLA Physics Case}}, \href{http://dx.doi.org/10.1088/1361-6633/ab28d6}{\emph{Rept. Prog. Phys.} {\bf 82} (2019) 116201}, [\href{http://arxiv.org/abs/1806.07396}{{\tt 1806.07396}}].

\bibitem{Bauer:2019vqk}
M.~Bauer, O.~Brandt, L.~Lee and C.~Ohm, \emph{{ANUBIS: Proposal to search for long-lived neutral particles in CERN service shafts}},  \href{http://arxiv.org/abs/1909.13022}{{\tt 1909.13022}}.

\bibitem{Gligorov:2017nwh}
V.~V. Gligorov, S.~Knapen, M.~Papucci and D.~J. Robinson, \emph{{Searching for Long-lived Particles: A Compact Detector for Exotics at LHCb}}, \href{http://dx.doi.org/10.1103/PhysRevD.97.015023}{\emph{Phys. Rev. D} {\bf 97} (2018) 015023}, [\href{http://arxiv.org/abs/1708.09395}{{\tt 1708.09395}}].

\bibitem{Gligorov:2018vkc}
V.~V. Gligorov, S.~Knapen, B.~Nachman, M.~Papucci and D.~J. Robinson, \emph{{Leveraging the ALICE/L3 cavern for long-lived particle searches}}, \href{http://dx.doi.org/10.1103/PhysRevD.99.015023}{\emph{Phys. Rev. D} {\bf 99} (2019) 015023}, [\href{http://arxiv.org/abs/1810.03636}{{\tt 1810.03636}}].

\bibitem{Tucker-Smith:2001myb}
D.~Tucker-Smith and N.~Weiner, \emph{{Inelastic dark matter}}, \href{http://dx.doi.org/10.1103/PhysRevD.64.043502}{\emph{Phys. Rev. D} {\bf 64} (2001) 043502}, [\href{http://arxiv.org/abs/hep-ph/0101138}{{\tt hep-ph/0101138}}].

\bibitem{Dienes:2011ja}
K.~R. Dienes and B.~Thomas, \emph{{Dynamical Dark Matter: I. Theoretical Overview}}, \href{http://dx.doi.org/10.1103/PhysRevD.85.083523}{\emph{Phys. Rev. D} {\bf 85} (2012) 083523}, [\href{http://arxiv.org/abs/1106.4546}{{\tt 1106.4546}}].

\bibitem{Dienes:2011sa}
K.~R. Dienes and B.~Thomas, \emph{{Dynamical Dark Matter: II. An Explicit Model}}, \href{http://dx.doi.org/10.1103/PhysRevD.85.083524}{\emph{Phys. Rev. D} {\bf 85} (2012) 083524}, [\href{http://arxiv.org/abs/1107.0721}{{\tt 1107.0721}}].

\bibitem{Dienes:2012jb}
K.~R. Dienes and B.~Thomas, \emph{{Phenomenological Constraints on Axion Models of Dynamical Dark Matter}}, \href{http://dx.doi.org/10.1103/PhysRevD.86.055013}{\emph{Phys. Rev. D} {\bf 86} (2012) 055013}, [\href{http://arxiv.org/abs/1203.1923}{{\tt 1203.1923}}].

\bibitem{Hochberg:2015vrg}
Y.~Hochberg, E.~Kuflik and H.~Murayama, \emph{{SIMP Spectroscopy}}, \href{http://dx.doi.org/10.1007/JHEP05(2016)090}{\emph{JHEP} {\bf 05} (2016) 090}, [\href{http://arxiv.org/abs/1512.07917}{{\tt 1512.07917}}].

\bibitem{Strassler:2006im}
M.~J. Strassler and K.~M. Zurek, \emph{{Echoes of a hidden valley at hadron colliders}}, \href{http://dx.doi.org/10.1016/j.physletb.2007.06.055}{\emph{Phys. Lett. B} {\bf 651} (2007) 374--379}, [\href{http://arxiv.org/abs/hep-ph/0604261}{{\tt hep-ph/0604261}}].

\bibitem{Strassler:2006ri}
M.~J. Strassler and K.~M. Zurek, \emph{{Discovering the Higgs through highly-displaced vertices}}, \href{http://dx.doi.org/10.1016/j.physletb.2008.02.008}{\emph{Phys. Lett. B} {\bf 661} (2008) 263--267}, [\href{http://arxiv.org/abs/hep-ph/0605193}{{\tt hep-ph/0605193}}].

\bibitem{Holdom:1985ag}
B.~Holdom, \emph{{Two U(1)'s and Epsilon Charge Shifts}}, \href{http://dx.doi.org/10.1016/0370-2693(86)91377-8}{\emph{Phys. Lett. B} {\bf 166} (1986) 196--198}.

\bibitem{Bauer:2018onh}
M.~Bauer, P.~Foldenauer and J.~Jaeckel, \emph{{Hunting All the Hidden Photons}}, \href{http://dx.doi.org/10.1007/JHEP07(2018)094}{\emph{JHEP} {\bf 07} (2018) 094}, [\href{http://arxiv.org/abs/1803.05466}{{\tt 1803.05466}}].

\bibitem{Fabbrichesi:2020wbt}
M.~Fabbrichesi, E.~Gabrielli and G.~Lanfranchi, \emph{{The Dark Photon}},  \href{http://arxiv.org/abs/2005.01515}{{\tt 2005.01515}}.

\bibitem{Caputo:2021eaa}
A.~Caputo, A.~J. Millar, C.~A.~J. O'Hare and E.~Vitagliano, \emph{{Dark photon limits: A handbook}}, \href{http://dx.doi.org/10.1103/PhysRevD.104.095029}{\emph{Phys. Rev. D} {\bf 104} (2021) 095029}, [\href{http://arxiv.org/abs/2105.04565}{{\tt 2105.04565}}].

\bibitem{Peccei:1977hh}
R.~D. Peccei and H.~R. Quinn, \emph{{CP Conservation in the Presence of Instantons}}, \href{http://dx.doi.org/10.1103/PhysRevLett.38.1440}{\emph{Phys. Rev. Lett.} {\bf 38} (1977) 1440--1443}.

\bibitem{Peccei:1977ur}
R.~D. Peccei and H.~R. Quinn, \emph{{Constraints Imposed by CP Conservation in the Presence of Instantons}}, \href{http://dx.doi.org/10.1103/PhysRevD.16.1791}{\emph{Phys. Rev. D} {\bf 16} (1977) 1791--1797}.

\bibitem{Jaeckel:2010ni}
J.~Jaeckel and A.~Ringwald, \emph{{The Low-Energy Frontier of Particle Physics}}, \href{http://dx.doi.org/10.1146/annurev.nucl.012809.104433}{\emph{Ann. Rev. Nucl. Part. Sci.} {\bf 60} (2010) 405--437}, [\href{http://arxiv.org/abs/1002.0329}{{\tt 1002.0329}}].

\bibitem{Bauer:2017ris}
M.~Bauer, M.~Neubert and A.~Thamm, \emph{{Collider Probes of Axion-Like Particles}}, \href{http://dx.doi.org/10.1007/JHEP12(2017)044}{\emph{JHEP} {\bf 12} (2017) 044}, [\href{http://arxiv.org/abs/1708.00443}{{\tt 1708.00443}}].

\bibitem{Gell-Mann:1979vob}
M.~Gell-Mann, P.~Ramond and R.~Slansky, \emph{{Complex Spinors and Unified Theories}}, {\emph{Conf. Proc. C} {\bf 790927} (1979) 315--321}, [\href{http://arxiv.org/abs/1306.4669}{{\tt 1306.4669}}].

\bibitem{Mohapatra:1979ia}
R.~N. Mohapatra and G.~Senjanovic, \emph{{Neutrino Mass and Spontaneous Parity Nonconservation}}, \href{http://dx.doi.org/10.1103/PhysRevLett.44.912}{\emph{Phys. Rev. Lett.} {\bf 44} (1980) 912}.

\bibitem{Schechter:1980gr}
J.~Schechter and J.~W.~F. Valle, \emph{{Neutrino Masses in SU(2) x U(1) Theories}}, \href{http://dx.doi.org/10.1103/PhysRevD.22.2227}{\emph{Phys. Rev. D} {\bf 22} (1980) 2227}.

\bibitem{Asaka:2005pn}
T.~Asaka and M.~Shaposhnikov, \emph{{The $\nu$MSM, dark matter and baryon asymmetry of the universe}}, \href{http://dx.doi.org/10.1016/j.physletb.2005.06.020}{\emph{Phys. Lett. B} {\bf 620} (2005) 17--26}, [\href{http://arxiv.org/abs/hep-ph/0505013}{{\tt hep-ph/0505013}}].

\bibitem{Kersten:2007vk}
J.~Kersten and A.~Y. Smirnov, \emph{{Right-Handed Neutrinos at CERN LHC and the Mechanism of Neutrino Mass Generation}}, \href{http://dx.doi.org/10.1103/PhysRevD.76.073005}{\emph{Phys. Rev. D} {\bf 76} (2007) 073005}, [\href{http://arxiv.org/abs/0705.3221}{{\tt 0705.3221}}].

\bibitem{Drewes:2015iva}
M.~Drewes and B.~Garbrecht, \emph{{Combining experimental and cosmological constraints on heavy neutrinos}}, \href{http://dx.doi.org/10.1016/j.nuclphysb.2017.05.001}{\emph{Nucl. Phys. B} {\bf 921} (2017) 250--315}, [\href{http://arxiv.org/abs/1502.00477}{{\tt 1502.00477}}].

\bibitem{Deppisch:2018eth}
F.~F. Deppisch, W.~Liu and M.~Mitra, \emph{{Long-lived Heavy Neutrinos from Higgs Decays}}, \href{http://dx.doi.org/10.1007/JHEP08(2018)181}{\emph{JHEP} {\bf 08} (2018) 181}, [\href{http://arxiv.org/abs/1804.04075}{{\tt 1804.04075}}].

\bibitem{Amrith:2018yfb}
S.~Amrith, J.~M. Butterworth, F.~F. Deppisch, W.~Liu, A.~Varma and D.~Yallup, \emph{{LHC Constraints on a $B-L$ Gauge Model using Contur}}, \href{http://dx.doi.org/10.1007/JHEP05(2019)154}{\emph{JHEP} {\bf 05} (2019) 154}, [\href{http://arxiv.org/abs/1811.11452}{{\tt 1811.11452}}].

\bibitem{Deppisch:2019kvs}
F.~Deppisch, S.~Kulkarni and W.~Liu, \emph{{Heavy neutrino production via $Z'$ at the lifetime frontier}}, \href{http://dx.doi.org/10.1103/PhysRevD.100.035005}{\emph{Phys. Rev. D} {\bf 100} (2019) 035005}, [\href{http://arxiv.org/abs/1905.11889}{{\tt 1905.11889}}].

\bibitem{Liu:2022kid}
W.~Liu, S.~Kulkarni and F.~F. Deppisch, \emph{{Heavy neutrinos at the FCC-hh in the U(1)B-L model}}, \href{http://dx.doi.org/10.1103/PhysRevD.105.095043}{\emph{Phys. Rev. D} {\bf 105} (2022) 095043}, [\href{http://arxiv.org/abs/2202.07310}{{\tt 2202.07310}}].

\bibitem{Liu:2022ugx}
W.~Liu, J.~Li, J.~Li and H.~Sun, \emph{{Testing the seesaw mechanisms via displaced right-handed neutrinos from a light scalar at the HL-LHC}}, \href{http://dx.doi.org/10.1103/PhysRevD.106.015019}{\emph{Phys. Rev. D} {\bf 106} (2022) 015019}, [\href{http://arxiv.org/abs/2204.03819}{{\tt 2204.03819}}].

\bibitem{Liu:2023nxi}
W.~Liu and Y.~Zhang, \emph{{Testing neutrino dipole portal by long-lived particle detectors at the LHC}}, \href{http://dx.doi.org/10.1140/epjc/s10052-023-11751-0}{\emph{Eur. Phys. J. C} {\bf 83} (2023) 568}, [\href{http://arxiv.org/abs/2302.02081}{{\tt 2302.02081}}].

\bibitem{Zhang:2023nxy}
Y.~Zhang and W.~Liu, \emph{{Probing active-sterile neutrino transition magnetic moments at LEP and CEPC}}, \href{http://dx.doi.org/10.1103/PhysRevD.107.095031}{\emph{Phys. Rev. D} {\bf 107} (2023) 095031}, [\href{http://arxiv.org/abs/2301.06050}{{\tt 2301.06050}}].

\bibitem{Barducci:2023hzo}
D.~Barducci, W.~Liu, A.~Titov, Z.~S. Wang and Y.~Zhang, \emph{{Probing the dipole portal to heavy neutral leptons via meson decays at the high-luminosity LHC}}, \href{http://dx.doi.org/10.1103/PhysRevD.108.115009}{\emph{Phys. Rev. D} {\bf 108} (2023) 115009}, [\href{http://arxiv.org/abs/2308.16608}{{\tt 2308.16608}}].

\bibitem{Deppisch:2023sga}
F.~F. Deppisch, S.~Kulkarni and W.~Liu, \emph{{Sterile Neutrinos at MAPP in the B-L Model}},  11, 2023.
\newblock \href{http://arxiv.org/abs/2311.01719}{{\tt 2311.01719}}.

\bibitem{Li:2023dbs}
J.~Li, W.~Liu and H.~Sun, \emph{{Z' Mediated right-handed Neutrinos from Meson Decays at the FASER}},  \href{http://arxiv.org/abs/2309.05020}{{\tt 2309.05020}}.

\bibitem{Liu:2023klu}
W.~Liu and F.~F. Deppisch, \emph{{Testing Leptogenesis and Seesaw using Long-lived Particle Searches in the $B-L$ Model}},  \href{http://arxiv.org/abs/2312.11165}{{\tt 2312.11165}}.

\bibitem{Bandyopadhyay:2023joz}
P.~Bandyopadhyay, M.~Frank, S.~Parashar and C.~Sen, \emph{{Interplay of inert doublet and vector-like lepton triplet with displaced vertices at the LHC/FCC and MATHUSLA}}, \href{http://dx.doi.org/10.1007/JHEP03(2024)109}{\emph{JHEP} {\bf 03} (2024) 109}, [\href{http://arxiv.org/abs/2310.08883}{{\tt 2310.08883}}].

\bibitem{Cao:2023smj}
Q.-H. Cao, J.~Guo, J.~Liu, Y.~Luo and X.-P. Wang, \emph{{Long-lived Searches of Vector-like Lepton and Its Accompanying Scalar at Colliders}},  \href{http://arxiv.org/abs/2311.12934}{{\tt 2311.12934}}.

\bibitem{Alimena:2019zri}
J.~Alimena et~al., \emph{{Searching for long-lived particles beyond the Standard Model at the Large Hadron Collider}}, \href{http://dx.doi.org/10.1088/1361-6471/ab4574}{\emph{J. Phys. G} {\bf 47} (2020) 090501}, [\href{http://arxiv.org/abs/1903.04497}{{\tt 1903.04497}}].

\bibitem{Kling:2022uzy}
F.~Kling, S.~Li, H.~Song, S.~Su and W.~Su, \emph{{Light Scalars at FASER}}, \href{http://dx.doi.org/10.1007/JHEP08(2023)001}{\emph{JHEP} {\bf 08} (2023) 001}, [\href{http://arxiv.org/abs/2212.06186}{{\tt 2212.06186}}].

\bibitem{Liu:2022nvk}
W.~Liu, A.~Yang and H.~Sun, \emph{{Shedding light on the electroweak phase transition from exotic Higgs boson decays at the lifetime frontiers}}, \href{http://dx.doi.org/10.1103/PhysRevD.105.115040}{\emph{Phys. Rev. D} {\bf 105} (2022) 115040}, [\href{http://arxiv.org/abs/2205.08205}{{\tt 2205.08205}}].

\bibitem{Haisch:2023rqs}
U.~Haisch and L.~Schnell, \emph{{Long-lived particle phenomenology in the 2HDM+a model}}, \href{http://dx.doi.org/10.1007/JHEP04(2023)134}{\emph{JHEP} {\bf 04} (2023) 134}, [\href{http://arxiv.org/abs/2302.02735}{{\tt 2302.02735}}].

\bibitem{Lee:1973iz}
T.~D. Lee, \emph{{A Theory of Spontaneous T Violation}}, \href{http://dx.doi.org/10.1103/PhysRevD.8.1226}{\emph{Phys. Rev. D} {\bf 8} (1973) 1226--1239}.

\bibitem{Branco:2011iw}
G.~C. Branco, P.~M. Ferreira, L.~Lavoura, M.~N. Rebelo, M.~Sher and J.~P. Silva, \emph{{Theory and phenomenology of two-Higgs-doublet models}}, \href{http://dx.doi.org/10.1016/j.physrep.2012.02.002}{\emph{Phys. Rept.} {\bf 516} (2012) 1--102}, [\href{http://arxiv.org/abs/1106.0034}{{\tt 1106.0034}}].

\bibitem{Wang:2022yhm}
L.~Wang, J.~M. Yang and Y.~Zhang, \emph{{Two-Higgs-doublet models in light of current experiments: a brief review}}, \href{http://dx.doi.org/10.1088/1572-9494/ac7fe9}{\emph{Commun. Theor. Phys.} {\bf 74} (2022) 097202}, [\href{http://arxiv.org/abs/2203.07244}{{\tt 2203.07244}}].

\bibitem{CMS:2016dhk}
{\scshape CMS} collaboration, V.~Khachatryan et~al., \emph{{Searches for invisible decays of the Higgs boson in pp collisions at $\sqrt{s}$ = 7, 8, and 13 TeV}}, \href{http://dx.doi.org/10.1007/JHEP02(2017)135}{\emph{JHEP} {\bf 02} (2017) 135}, [\href{http://arxiv.org/abs/1610.09218}{{\tt 1610.09218}}].

\bibitem{ATLAS:2015gvj}
{\scshape ATLAS} collaboration, G.~Aad et~al., \emph{{Search for invisible decays of a Higgs boson using vector-boson fusion in $pp$ collisions at $\sqrt{s}=8$ TeV with the ATLAS detector}}, \href{http://dx.doi.org/10.1007/JHEP01(2016)172}{\emph{JHEP} {\bf 01} (2016) 172}, [\href{http://arxiv.org/abs/1508.07869}{{\tt 1508.07869}}].

\bibitem{ATLAS:2017nyv}
{\scshape ATLAS} collaboration, M.~Aaboud et~al., \emph{{Search for an invisibly decaying Higgs boson or dark matter candidates produced in association with a $Z$ boson in $pp$ collisions at $\sqrt{s} =$ 13 TeV with the ATLAS detector}}, \href{http://dx.doi.org/10.1016/j.physletb.2017.11.049}{\emph{Phys. Lett. B} {\bf 776} (2018) 318--337}, [\href{http://arxiv.org/abs/1708.09624}{{\tt 1708.09624}}].

\bibitem{ParticleDataGroup:2020ssz}
{\scshape Particle Data Group} collaboration, P.~A. Zyla et~al., \emph{{Review of Particle Physics}}, \href{http://dx.doi.org/10.1093/ptep/ptaa104}{\emph{PTEP} {\bf 2020} (2020) 083C01}.

\bibitem{Wang:2013sha}
L.~Wang and X.-F. Han, \emph{{Status of the aligned two-Higgs-doublet model confronted with the Higgs data}}, \href{http://dx.doi.org/10.1007/JHEP04(2014)128}{\emph{JHEP} {\bf 04} (2014) 128}, [\href{http://arxiv.org/abs/1312.4759}{{\tt 1312.4759}}].

\bibitem{Deshpande:1977rw}
N.~G. Deshpande and E.~Ma, \emph{{Pattern of Symmetry Breaking with Two Higgs Doublets}}, \href{http://dx.doi.org/10.1103/PhysRevD.18.2574}{\emph{Phys. Rev. D} {\bf 18} (1978) 2574}.

\bibitem{Kanemura:1993hm}
S.~Kanemura, T.~Kubota and E.~Takasugi, \emph{{Lee-Quigg-Thacker bounds for Higgs boson masses in a two doublet model}}, \href{http://dx.doi.org/10.1016/0370-2693(93)91205-2}{\emph{Phys. Lett. B} {\bf 313} (1993) 155--160}, [\href{http://arxiv.org/abs/hep-ph/9303263}{{\tt hep-ph/9303263}}].

\bibitem{Akeroyd:2000wc}
A.~G. Akeroyd, A.~Arhrib and E.-M. Naimi, \emph{{Note on tree level unitarity in the general two Higgs doublet model}}, \href{http://dx.doi.org/10.1016/S0370-2693(00)00962-X}{\emph{Phys. Lett. B} {\bf 490} (2000) 119--124}, [\href{http://arxiv.org/abs/hep-ph/0006035}{{\tt hep-ph/0006035}}].

\bibitem{Eriksson:2009ws}
D.~Eriksson, J.~Rathsman and O.~Stal, \emph{{2HDMC: Two-Higgs-Doublet Model Calculator Physics and Manual}}, \href{http://dx.doi.org/10.1016/j.cpc.2009.09.011}{\emph{Comput. Phys. Commun.} {\bf 181} (2010) 189--205}, [\href{http://arxiv.org/abs/0902.0851}{{\tt 0902.0851}}].

\bibitem{Turner:1987by}
M.~S. Turner, \emph{{Axions from SN 1987a}}, \href{http://dx.doi.org/10.1103/PhysRevLett.60.1797}{\emph{Phys. Rev. Lett.} {\bf 60} (1988) 1797}.

\bibitem{Ellis:1987pk}
J.~R. Ellis and K.~A. Olive, \emph{{Constraints on Light Particles From Supernova Sn1987a}}, \href{http://dx.doi.org/10.1016/0370-2693(87)91710-2}{\emph{Phys. Lett. B} {\bf 193} (1987) 525}.

\bibitem{CHARM:1985anb}
{\scshape CHARM} collaboration, F.~Bergsma et~al., \emph{{Search for Axion Like Particle Production in 400-{GeV} Proton - Copper Interactions}}, \href{http://dx.doi.org/10.1016/0370-2693(85)90400-9}{\emph{Phys. Lett. B} {\bf 157} (1985) 458--462}.

\bibitem{Gorbunov:2021ccu}
D.~Gorbunov, I.~Krasnov and S.~Suvorov, \emph{{Constraints on light scalars from PS191 results}}, \href{http://dx.doi.org/10.1016/j.physletb.2021.136524}{\emph{Phys. Lett. B} {\bf 820} (2021) 136524}, [\href{http://arxiv.org/abs/2105.11102}{{\tt 2105.11102}}].

\bibitem{LHCb:2015nkv}
{\scshape LHCb} collaboration, R.~Aaij et~al., \emph{{Search for hidden-sector bosons in $B^0 \!\to K^{*0}\mu^+\mu^-$ decays}}, \href{http://dx.doi.org/10.1103/PhysRevLett.115.161802}{\emph{Phys. Rev. Lett.} {\bf 115} (2015) 161802}, [\href{http://arxiv.org/abs/1508.04094}{{\tt 1508.04094}}].

\bibitem{LHCb:2016awg}
{\scshape LHCb} collaboration, R.~Aaij et~al., \emph{{Search for long-lived scalar particles in $B^+ \to K^+ \chi (\mu^+\mu^-)$ decays}}, \href{http://dx.doi.org/10.1103/PhysRevD.95.071101}{\emph{Phys. Rev. D} {\bf 95} (2017) 071101}, [\href{http://arxiv.org/abs/1612.07818}{{\tt 1612.07818}}].

\bibitem{LHCb:2020car}
{\scshape LHCb} collaboration, R.~Aaij et~al., \emph{{Searches for 25 rare and forbidden decays of $D^{+}$ and $ {D}_s^{+} $ mesons}}, \href{http://dx.doi.org/10.1007/JHEP06(2021)044}{\emph{JHEP} {\bf 06} (2021) 044}, [\href{http://arxiv.org/abs/2011.00217}{{\tt 2011.00217}}].

\bibitem{NA62:2021zjw}
{\scshape NA62} collaboration, E.~Cortina~Gil et~al., \emph{{Measurement of the very rare K$^{+}$\textrightarrow{}$ {\pi}^{+}\nu \overline{\nu} $ decay}}, \href{http://dx.doi.org/10.1007/JHEP06(2021)093}{\emph{JHEP} {\bf 06} (2021) 093}, [\href{http://arxiv.org/abs/2103.15389}{{\tt 2103.15389}}].

\bibitem{MicroBooNE:2021usw}
{\scshape MicroBooNE} collaboration, P.~Abratenko et~al., \emph{{Search for a Higgs Portal Scalar Decaying to Electron-Positron Pairs in the MicroBooNE Detector}}, \href{http://dx.doi.org/10.1103/PhysRevLett.127.151803}{\emph{Phys. Rev. Lett.} {\bf 127} (2021) 151803}, [\href{http://arxiv.org/abs/2106.00568}{{\tt 2106.00568}}].

\bibitem{BNL-E949:2009dza}
{\scshape BNL-E949} collaboration, A.~V. Artamonov et~al., \emph{{Study of the decay $K^+\to\pi^+\nu \bar\nu$ in the momentum region $140 < P_\pi < 199$ MeV/c}}, \href{http://dx.doi.org/10.1103/PhysRevD.79.092004}{\emph{Phys. Rev. D} {\bf 79} (2009) 092004}, [\href{http://arxiv.org/abs/0903.0030}{{\tt 0903.0030}}].

\bibitem{Domingo:2016yih}
F.~Domingo, \emph{{Decays of a NMSSM CP-odd Higgs in the low-mass region}}, \href{http://dx.doi.org/10.1007/JHEP03(2017)052}{\emph{JHEP} {\bf 03} (2017) 052}, [\href{http://arxiv.org/abs/1612.06538}{{\tt 1612.06538}}].

\bibitem{Holstein:2001bt}
B.~R. Holstein, \emph{{Allowed eta decay modes and chiral symmetry}}, \href{http://dx.doi.org/10.1238/Physica.Topical.099a00055}{\emph{Phys. Scripta T} {\bf 99} (2002) 55--67}, [\href{http://arxiv.org/abs/hep-ph/0112150}{{\tt hep-ph/0112150}}].

\bibitem{ATLAS:2018nda}
{\scshape ATLAS} collaboration, M.~Aaboud et~al., \emph{{Search for dark matter in events with a hadronically decaying vector boson and missing transverse momentum in $pp$ collisions at $\sqrt{s} = 13$ TeV with the ATLAS detector}}, \href{http://dx.doi.org/10.1007/JHEP10(2018)180}{\emph{JHEP} {\bf 10} (2018) 180}, [\href{http://arxiv.org/abs/1807.11471}{{\tt 1807.11471}}].

\bibitem{MoEDAL-MAPP:2022kyr}
{\scshape MoEDAL-MAPP} collaboration, B.~Acharya et~al., \emph{{MoEDAL-MAPP, an LHC Dedicated Detector Search Facility}},  in \emph{{Snowmass 2021}}, 9, 2022.
\newblock \href{http://arxiv.org/abs/2209.03988}{{\tt 2209.03988}}.

\bibitem{MATHUSLA:2019qpy}
{\scshape MATHUSLA} collaboration, H.~Lubatti et~al., \emph{{Explore the lifetime frontier with MATHUSLA}}, \href{http://dx.doi.org/10.1088/1748-0221/15/06/C06026}{\emph{JINST} {\bf 15} (2020) C06026}, [\href{http://arxiv.org/abs/1901.04040}{{\tt 1901.04040}}].

\bibitem{Alloul:2013bka}
A.~Alloul, N.~D. Christensen, C.~Degrande, C.~Duhr and B.~Fuks, \emph{{FeynRules 2.0 - A complete toolbox for tree-level phenomenology}}, \href{http://dx.doi.org/10.1016/j.cpc.2014.04.012}{\emph{Comput. Phys. Commun.} {\bf 185} (2014) 2250--2300}, [\href{http://arxiv.org/abs/1310.1921}{{\tt 1310.1921}}].

\bibitem{Degrande:2011ua}
C.~Degrande, C.~Duhr, B.~Fuks, D.~Grellscheid, O.~Mattelaer and T.~Reiter, \emph{{UFO - The Universal FeynRules Output}}, \href{http://dx.doi.org/10.1016/j.cpc.2012.01.022}{\emph{Comput. Phys. Commun.} {\bf 183} (2012) 1201--1214}, [\href{http://arxiv.org/abs/1108.2040}{{\tt 1108.2040}}].

\bibitem{Alwall:2014hca}
J.~Alwall, R.~Frederix, S.~Frixione, V.~Hirschi, F.~Maltoni, O.~Mattelaer et~al., \emph{{The automated computation of tree-level and next-to-leading order differential cross sections, and their matching to parton shower simulations}}, \href{http://dx.doi.org/10.1007/JHEP07(2014)079}{\emph{JHEP} {\bf 07} (2014) 079}, [\href{http://arxiv.org/abs/1405.0301}{{\tt 1405.0301}}].

\bibitem{Sjostrand:2014zea}
T.~Sj\"ostrand, S.~Ask, J.~R. Christiansen, R.~Corke, N.~Desai, P.~Ilten et~al., \emph{{An introduction to PYTHIA 8.2}}, \href{http://dx.doi.org/10.1016/j.cpc.2015.01.024}{\emph{Comput. Phys. Commun.} {\bf 191} (2015) 159--177}, [\href{http://arxiv.org/abs/1410.3012}{{\tt 1410.3012}}].

\bibitem{MATHUSLA:2018bqv}
{\scshape MATHUSLA} collaboration, C.~Alpigiani et~al., \emph{{A Letter of Intent for MATHUSLA: A Dedicated Displaced Vertex Detector above ATLAS or CMS.}},  \href{http://arxiv.org/abs/1811.00927}{{\tt 1811.00927}}.

\bibitem{LHCHiggsWG}
``Lhc higgs cross section working group.'' \url{https://twiki.cern.ch/twiki/bin/view/LHCPhysics/HiggsEuropeanStrategy}.

\end{thebibliography}\endgroup
\end{document}